\documentclass[useAMS,usenatbib,usegraphicx,referee,a4paper]{mn2e}
\usepackage{graphicx}
\usepackage{lscape}
\title[Wide binarity and planet occurrence in the {\it Kepler} field]{A Pan-STARRS\,1 study of the relationship between wide binarity and planet occurrence in the {\it Kepler} field}
 \author[N.R.\ Deacon et al.]{N.R.\ Deacon\thanks{E-mail:n.deacon2@herts.ac.uk}$^{1,2}$, A.L.\ Kraus$^3$, A.W.\ Mann $^{4,5}$, E.A.\ Magnier$^6$, K.C.\ Chambers$^6$, \newauthor R.J.\ Wainscoat$^6$, J.L.\ Tonry$^6$, N.\ Kaiser$^6$, C.\ Waters$^6$, H.\ Flewelling$^6$,  \newauthor K.W.\ Hodapp$^7$, W.S.\ Burgett$^8$\\
$^1$Centre for Astrophysics Research, University of Hertfordshire, College Lane, Hatfield, AL1 5TL, UK\\
$^2$Max Planck Institute for Astronomy, Koenigstuhl 17, D-69117 Heidelberg, Germany\\
$^3$Department of Astronomy, The University of Texas at Austin, Austin, TX 78712, USA\\
$^4$Harlan J. Smith Fellow, Department of Astronomy, The University of Texas at Austin, Austin, TX 78712, USA\\
$^5$Visiting Researcher, Institute for Astrophysical Research, Boston University\\
$^6$Institute for Astronomy, University of Hawai'i, 2680 Woodlawn Drive, Honolulu, HI 96822, USA\\
$^7$Institute for Astronomy, University of Hawai'i, 640 North Aohoku Place, Hilo, HI 96720, USA\\
$^8$Giant Magellan Telescope Observatory, USA}

\begin{document}
 \date{}
 \pagerange{\pageref{firstpage}--\pageref{lastpage}} \pubyear{2015}
 \maketitle
 \label{firstpage}
 
 \begin{abstract}
 The NASA {\it Kepler} mission has revolutionised time-domain astronomy and has massively expanded the number of known extrasolar planets. However, the effect of wide multiplicity on exoplanet occurrence has not been tested with this dataset. We present a sample of 401 wide multiple systems containing at least one {\it Kepler} target star. Our method uses Pan-STARRS\,1 and archival data to produce an accurate proper motion catalogue of the {\it Kepler} field. Combined with Pan-STARRS\,1 SED fits and archival proper motions for bright stars, we use a newly developed probabilistic algorithm to identify likely wide binary pairs which are not chance associations. As by-products of this we present stellar SED templates in the Pan-STARRS\,1 photometric system and conversions from this system to {\it Kepler} magnitudes. We find that {\it Kepler} target stars in our binary sample with separations above 6$\arcsec$ are no more or less likely to be identified as confirmed or candidate planet hosts than a weighted comparison sample of {\it Kepler} targets of similar brightness and spectral type. Therefore we find no evidence that binaries with projected separations greater than 3,000\,AU affect the occurrence rate of planets with P$<$300\,days around FGK stars.

 \end{abstract}
% \slugcomment{Last Compiled \today}
 \section{Introduction}
The {\it Kepler} Mission \citep{Borucki2010} has to date identified 3697 exoplanet candidates (\citealt{Rowe2015} and references therein). The confirmed candidates and inferred false positive rates have been used to determine the faction of stars with shorter period planets (\citealt{Howard2011};\citealt{Fressin2013}) with particular attention paid to the number of Earth-like planets around M dwarfs (\citealt{Morton2014};\citealt{Dressing2015}). The determination of the parameters of transits exoplanets dependents heavily on the observed properties of the host star. Hence the study of exoplanet hosts' metallicity, radius and temperature has been a fruitful endeavour for several groups \citep{Mann2012,Mann2013b,Mann2013a,Muirhead2014,Newton2015}. One key problem with late-type stars is the significant pollution of the dwarf sequence by background giants. While these can be readily identified through spectroscopy \citep{Mann2012} this is hard to do for large numbers of stars. Because planets are difficult to impossible to find around giant stars using transit searches (and because evolved stars can cause errors in planet property determinations, \citealt{Bastien2014,Wang2015}), planet occurrence calculations require an estimate of the number of evolved stars in the survey \citep{Mann2012}. Therefore, calculating the planet fraction requires an accurate characterisation of the tens of thousands of {\it Kepler} target stars with no detected planet. \cite{Mann2013a} showed that reduced proper motion is an efficient dwarf/giant discriminator for late~K and M~dwarfs which can be used on much larger samples of stars. 

The quality of proper motions in the {\it Kepler} field varies with the magnitude of the target star. For the brightest objects either Hipparcos (to $V \approx 9$) and Tycho (to $V \approx 11.5$) provide proper motion measurements with uncertainties of the order of a few millarcseconds per year. Astrographic observations in the UCAC~4 catalogue provide proper motion errors below 10 milliarcseconds per year for stars down to $R\approx16$. This covers the majority of the {\it Kepler} targets but will miss any faint common proper motion companions to these {\it Kepler} targets. Digitised archival photometric surveys such as SuperCOSMOS \citep{Hambly2001} and USNO-B \citep{Monet2003} provide proper motions for these fainter stars. While these surveys have the advantage of long time baselines, their photographic nature means that each individual epoch is of poorer astrometric quality than modern CCD surveys. Additionally the most extensive proper motion survey based on photographic plates, the Superblink survey (with errors of approximately 8 milliarcseconds per year) only has a  few subsets of data which are publicly available: high proper motion stars in the Northern hemisphere \citep{Lepine2005} and bright M~dwarfs \citep{Lepine2011}.

Wide stellar companions are common in the field, with $>$25\% of nearby solar type stars having companions wider than 100\,AU \citep{Raghavan2010}. Among young stars, disk frequency \citep{Kraus2012} and disk mass \citep{Pascucci2007,Harris2012} are suppressed for close binaries but the components of systems wider than 100\,AU have similar disk properties to single stars. This suggests that wide binaries play a minor role in the initial sculpting of planetary systems. \cite{Raghavan2006} found 24 companions wider than 100\,AU amongst a sample of 131 radial velocity exoplanet hosts. This equates to a wide companion fraction of 18.3$\pm$3.7\%, slightly lower than the $\sim$25\% of solar-type stars in \cite{Raghavan2010} which have companions wider than 100\,AU. Finally, \cite{Wang2014} studied binary systems closer than 1500\,AU finding a significant deficit of planets around the closest binaries with the effect dropping significantly becoming very small for systems wider than 100\,AU.

Recent work by \cite{Kaib2013} suggests a different mechanism by which binaries with semi major axes of 1,000\,AU or so can affect planetary systems. While orbiting through the Milky Way's gravitational potential, the orbital parameters of wide binary companions can be modified by gravitational interactions with passing stars driving them to more elliptical orbits and sending the companions close to their primary at periastron. Such close passage can disrupt the outer planets in such systems and induce planetary migration and/or ejection. The magnitude of this effect can be tested by comparing the planet occurrence rate around wide binary components with that around field stars. However this is complicated by coincident alignments of unrelated pairs of stars often mimicking true wide binaries. Accurate kinematic measurements and statistical models are required to separate true wide binaries from coincident pairings.

Wide binaries are also useful astrophysical probes, ideal for calibrating and testing empirical stellar property estimators. Gyrochronology, the process of estimating a star's age from its rotation period, is one of the most valuable techniques in stellar astronomy \citep{Barnes2007}. This can be used to derive ages for exoplanet host stars (e.g. \citealt{Walkowicz2013}) or ultracool companions to intermediate mass stars (e.g. \citealt{Dupuy2008}). These relations are often calibrated using open clusters of stars as anchor points. Each cluster provides a few hundred stars of different types, all of the same age. These discrete anchor points have the disadvantage that there are relatively few well studied, older open clusters with well determined rotation periods for their stellar members. \cite{Meibom2015} recently provided rotation periods for the 2.5\,Gyr-old cluster NGC 6819. More field-age tests of gyrochronology relations will improve stellar age estimates even further, but each cluster still only represents a single age. Wide binary systems, however, cover a wide range of ages. While their age is not known independently as with clusters, they can still be used as a powerful test of gyrochonology at ages outside the range of known clusters. Wide binary systems are coeval systems and thus their components should have similar ages measured from gyrochronology. {\it Kepler} provides excellent photometry for rotation period calculation so any wide binaries of two {\it Kepler} target stars can be used to test gyrochronology relations. Conversely, gyrochronology relations can be used to test if wide binary systems are true physical pairs.

In this paper we present a proper motion survey of the {\it Kepler} field. We combine archival datasets from wide-field public surveys and target UKIRT observations with new Pan-STARRS\,1 astrometry to produce accurate proper motions for {\it Kepler} target stars. This enables us to combine the excellent astrometric accuracy of a modern CCD-based survey with the long time baseline of archival plate data. We then use our Pan-STARRS\,1 plus archival data proper motion catalogue, along with SED fits using Pan-STARRS\,1 and 2MASS photometry, to select a population of wide binary pairs where at least one component is a {\it Kepler} target star. We then use these binaries to test whether wide binarity has an effect on exoplanet occurrence.

\section{Datasets}
\subsection{Pan-STARRS\,1 data}
A 1.8\,m high etendue survey telescope, Pan-STARRS\,1 \citep{Kaiser2002} recently completed its full three and a half year survey operations on Haleakala on Maui in the Hawaiian Islands. This consisted of a suite of surveys of different cadences and depths. The {\it Kepler} field was included in the 3$\pi$ survey which covers the full sky north of $\delta=-30^{\circ}$ in five filters ($g_{P1}$, $r_{P1}$, $i_{P1}$, $z_{P1}$ and $y_{P1}$; \citealt{Tonry2012}). Each filter has approximately six pairs of observations each separated by roughly half an hour over the course of the survey for any one point on the sky, therefore a typical object will have 30 pairs of observations. The data from Pan-STARRS\,1 are astrometrically and photometrically reduced and calibrated using the processes outlined in \cite{Magnier2006}, \cite{Magnier2007}, \cite{Schlafly2012} and \cite{Magnier2013}.
\subsubsection{Astrometric accuracy}
There are three separate uncertainties in our Pan-STARRS\,1 astrometric uncertainty model: centroiding errors caused by photon noise, the Pan-STARRS\,1 internal systematic floor and the systematic floor caused by comparing Pan-STARRS\,1 with other surveys.

\begin{figure}
\begin{center}
\begin{picture}(100,200)
 \includegraphics{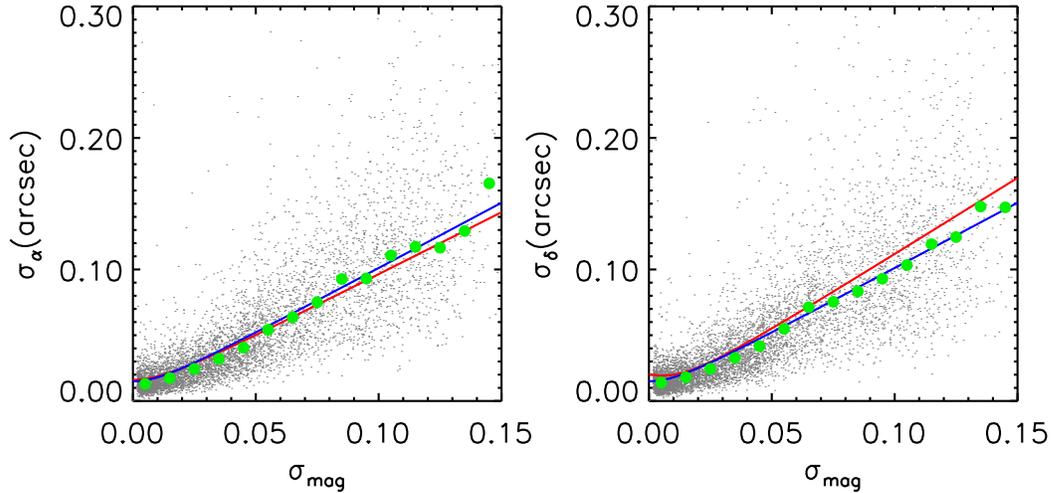}
 \end{picture}
\caption{\label{ps1_ps1} Astrometric scatter of $r_{P1}$ detections about mean object positions as a function of mean photometric error for the object's $r_{P1}$ detections. The red line shows a least squares fit of the form $\sigma_{pos} = \sqrt{a^2 + b^2 \sigma_{mag}^2}$ and the blue line shows our assumed astrometric model with $a$=15mas and $b$=1. Green dots are the median scatter in each 0.01\,mag bin.}
\end{center}
\end{figure}

The astrometric centroiding error typically takes the form $\sigma_{pos}=\sqrt{a^2 + b^2\sigma_{mag}^2}$ where $\sigma_{pos}$ is our positional error in arcseconds, $\sigma_{mag}$ is the photometric error in magnitudes, $a$ is our systematic floor and $b$ is some conversion factor which depends on the PSF shape and whether the object lies in the source or background dominated noise regime \citep{King1983}. The Pan-STARRS\,1 PSF takes a form which includes a free parameter (Magnier et al., in prep.). Thus we cannot know $b$ without knowing the exact PSF shape for every object. As this would be cumbersome and computationally impractical to implement, we take an empirical approach. We extracted all $r_{P1}$ detections in a 1$\times$1 degree patch centred on $R.A.= 18^h 42^m$ $Dec. = 46^{\circ} 30^m$ and measured the scatter in each object's position about the mean position. We then plotted this as a function of the mean photometric error for the detections of that object (see Figure~\ref{ps1_ps1}). We found that a model of the form $\sigma_{pos} = \sqrt{0.015^2 + \sigma_{mag}^2}$ was a good fit to the data. Hence we adopt 15\,milliarcseconds as our Pan-STARRS\,1 internal systematic floor and set our photon-noise positional errors to be the same as the magnitude error (i.e. $b=1$). 

To estimate the systematic error added by comparing Pan-STARRS\,1 data with other surveys we again took another 1$\times$1 degree test area (this time centred on $R.A.= 19^h 14^m$ $Dec. = 38^{\circ} 30^m$ as our first field has no SDSS data) and examined the positional scatter between Pan-STARRS\,1 and three external surveys (SDSS, \citealt{SDSS8}; 2MASS, \citealt{Skrutskie2006} and P.I. UKIRT data of the {\it Kepler} field processed using the same pipeline as UKIDSS data, \citealt{Lawrence2007}). Selecting only bright objects which lay in the magnitude range of each of the surveys where the astrometric uncertainty is dominated by systematic errors ($15<r<16$ for SDSS and Pan-STARRS\,1, $11<J<15$ for 2MASS and UKIRT) we examined what the scatter about zero was for each pair of surveys. This provided us with six linear combinations of the systematic variances of the surveys. We solved these as a set of equations by singular value decomposition finding that the systematic variance of Pan-STARRS\,1 data compared to external datasets is approximately 38\,milliarcseconds (including our internal systematic floor of 15 milliarcseconds). Once the internal systematic floor is subtracted in quadrature, we find a systematic error going from the Pan-STARRS\,1 reference frame to some external reference of 35\,milliarcseconds. Both external floor and internal floor are used later in our calculation of proper motions.

\subsection{UKIRT data}
We made use of observations which were made of the {\it Kepler} field using WFCAM \citep{Casali2007} on the UK Infrared telescope which are now public. These data came from program U/09A/2 (PI Lucas) observed on July 11th--13th 2009. 
Observations were done in the $J_{MKO}$-band in a similar set-up the the UKIDSS Large Area Survey $J_{MKO}$ observations (2$\times$2 microstepping, one telescope offset, 40\,s total integration, \citealt{Lawrence2007}) with reduced data catalogues available as FITS tables in the WFCAM Science Archive \citep{Hambly2008} having been processed by the WFCAM pipeline \citep{Irwin2004}. 

\subsection{Public survey data}
We included astrometry from multiple public surveys in our work. We drew data from the Two Micron All Sky Survey (2MASS; \citealt{Skrutskie2006}), the Sloan Digital Sky Survey (SDSS DR8;  \citealt{SDSS8}), the USNO-B digitisation of photographic plate data \citep{Monet2003} and the WISE All-Sky Survey \citep{Wright2010}. The astrometry and photometry from these surveys were all extracted from the {\it Vizier} online archive service with the exception of WISE dataset which was downloaded from the NASA/IPAC Infrared Science Archive. To avoid spurious faint photographic plate detections affecting our proper motion fits we ignored any detections in USNO-B fainter than $B_J=19$, $R$=19 and $I_N$=18. This last cut is similar to that used in \cite{Deacon2005} and \cite{Deacon2007}.

For archival surveys we used the astrometric error estimates provided by the relevant survey archives with the exception of SDSS. For this survey we used the error model used by \cite{Kraus2007} which features a 40 milliarcsecond systematic floor (similar to that found by \citealt{Pier2002}) and photon noise term that scales as the quoted magnitude uncertainty.

\section{Calculation of proper motions}
We divided the {\it {\it Kepler}} field into 1$\times$1 degrees chunks and in each of these areas extracted Pan-STARRS\,1 data from the second large-scale reprocessing of the data (PV2) using scripts written in the {\it Desktop Virtual Observatory} shell language \citep{Magnier2008}. We extracted as our base catalogue the average catalog objects in the target area. This average object catalogue consists of photometric and astrometric properties derived from all observations of each individual object  and includes initial Pan-STARRS\,1-only proper motions. As some objects will have proper motions that will take their archival detections outside our pairing radius of 1$\arcsec$, we used the calculated Pan-STARRS\,1-only proper motions to find the likely position of each object at the epochs of each archival non-Pan-STARRS observation. We then searched around these positions with a pairing radius of 1\,$\arcsec$ to identify the objects' detection in each survey. This procedure is vulnerable to spurious proper motions in the Pan-STARRS\,1 database so we do not applying the proper motion ``rewind" procedure for any object with a Pan-STARRS\,1 proper motion $\chi^2_{\nu}>10$. 

To calculate our proper motions we constructed a covariance matrix taking into account that the Pan-STARRS\,1 measurements will have three errors associated with them, photon noise, the systematic floor of relative Pan-STARRS\,1 astrometry and the systematic error between Pan-STARRS\,1 and other surveys. The Pan-STARRS\,1 measurements will have errors correlated with each other with a covariance of the square of the Pan-STARRS\,1 to other surveys systematic floor. No other survey data was given off-diagonal terms in the covariance matrix. We then calculated the proper motion for each object based on the measurements and our covariance matrix.

\begin{figure}
\begin{center}
\includegraphics[scale=1.0]{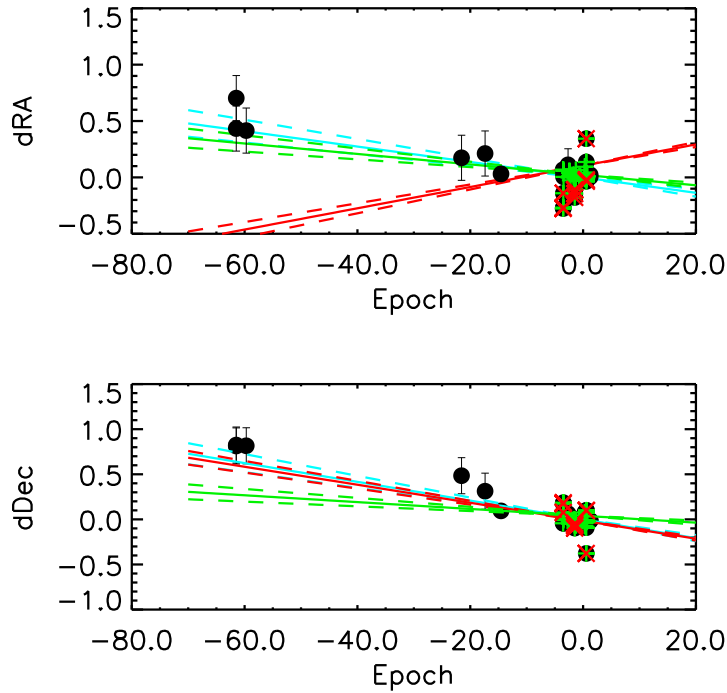}
\caption{\label{ex_clipping} An example of the fitting and clipping process for the star KIC 6878408. The initial fit is the red line, the second fit after the 10$\sigma$ clip is the green line and the final fit after the 3$\sigma$ clip is the blue line. The dashed lines represent the 1$\sigma$ confidence limits on the proper motion in each fit. Note points crossed out with red are excluded by the 10$\sigma$ clip and those crossed out in green are excluded by the 3 $\sigma$ clip.}
\end{center}
\end{figure}

We performed our fit in three stages. Having examined a number of preliminary fits we found that these were often offset by one or two Pan-STARRS data points which were significant outliers. Performing a 3$\sigma$ clip around our initial fit would be the most obvious solution. However this initial fit would be strongly affected by the outlier points and hence may exclude many valid datapoints which were outliers to this erroneous initial proper motion fit. Hence we performed an initial clipping removing any Pan-STARRS\,1 data point which was more than a 10$\sigma$ outlier (i.e. a 10$\sigma$ outlier in the quadrature sum of the number of standard deviations the object was an outlier from the fit in R.A. and Dec.). After this, we refitted our data and applied a more stringent 3$\sigma$ cut again only on the Pan-STARRS\,1 data. This was followed by a final refitting of our proper motions. Figure~\ref{ex_clipping} shows and example of the clipping and fitting process on one {\it Kepler} target.

\subsection{Selecting the proper motions of Kepler targets}
Often there are multiple potential Pan-STARRS\,1 matches for a particular Kepler target. We followed the following procedure to select the most appropriate matches. Firstly we selected only objects which had $g_{P1}$, $r_{P1}$ or $i_{P1}$ magnitudes within 2 magnitudes of the $g$, $r$ and $i$ magnitudes of the Kepler target in the Kepler Input Catalog. This excludes matches with obviously spurious faint sources. We then matched each Kepler target with the  predicted Epoch=1999.0 positions of all objects within 3$\arcsec$ and implemented the following procedure for multiple matches.
\begin{enumerate} 
	\item We asked how many of those matches are within $1\arcsec$ and have $r$ magnitudes in the KIC that agree with Pan-STARRS\,1 $r_{P1}$ to within 0.2\,mag.
	\begin{enumerate}
		\item If there is one such match we use this one
		\item If there are multiple of these we use the one with the most measurements in its astrometric fit
		\end{enumerate} 
	\item If not  we ask how many of the matches have $r$ magnitudes in the KIC that agree with Pan-STARRS\,1 $r_{P1}$ to within 0.2mags
	\begin{enumerate}
		\item If there is one such match we use this one
		\item If there are multiple of these we use the closest one
		\end{enumerate} 
	\item Otherwise we use the closest positional match
\end{enumerate}

\subsection{Proper motion results}
\begin{figure}
\begin{center}
\begin{tabular}{cc}
\includegraphics[scale=0.4]{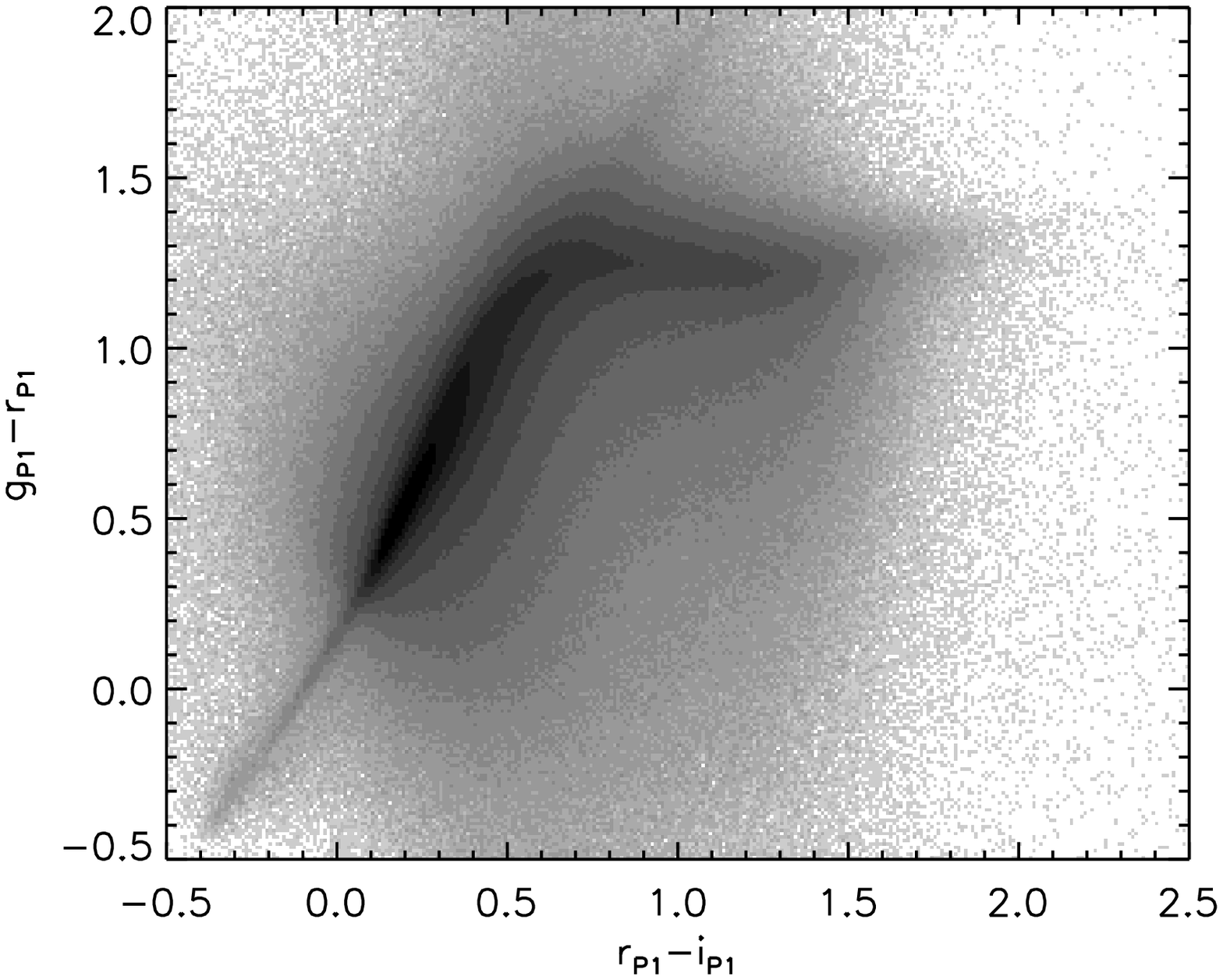}&\includegraphics[scale=0.4]{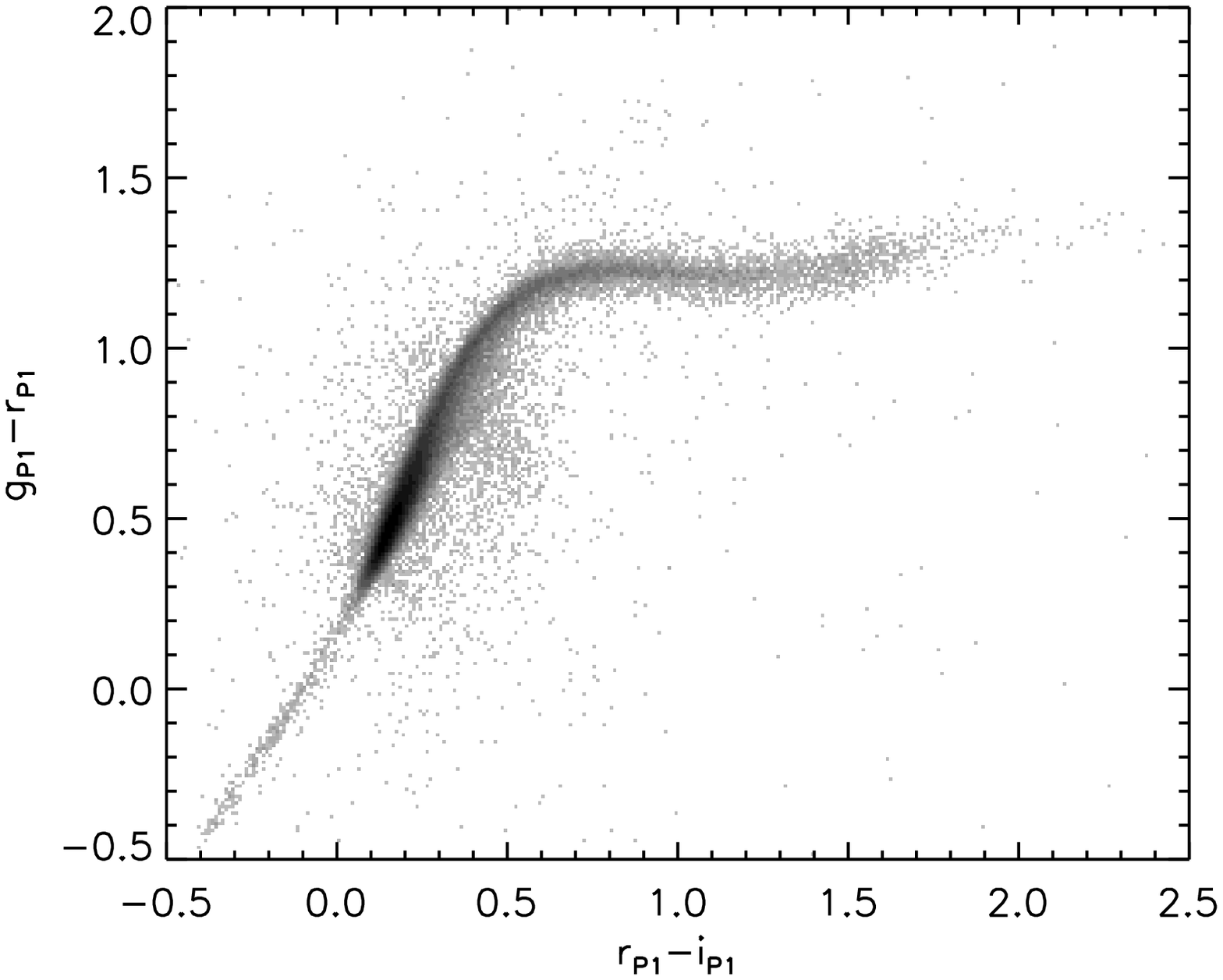}
\end{tabular}
\caption{\label{ps1_ccd} The $\log_{10}$ density distribution of all unsaturated ($r_{P1}>14.5$\,mag.) sources (left) and unsaturated {\it Kepler} targets (right) on a colour-colour plot. Note the concentration of {\it Kepler} targets around (0.2,0.5) in the {\it Kepler} targets plot. This is due to the large number of F and G stars in the {\it Kepler} target list. Note also the lack of an extension of the F, G, K locus after the stellar locus takes a turn at (0.5,1.2) with the onset of TiO absorption in M dwarfs. This suggests few {\it Kepler} targets in our sample are significantly reddened.}
\end{center}
\end{figure}

\begin{figure}
\begin{center}
\begin{tabular}{cc}
\includegraphics[scale=0.5]{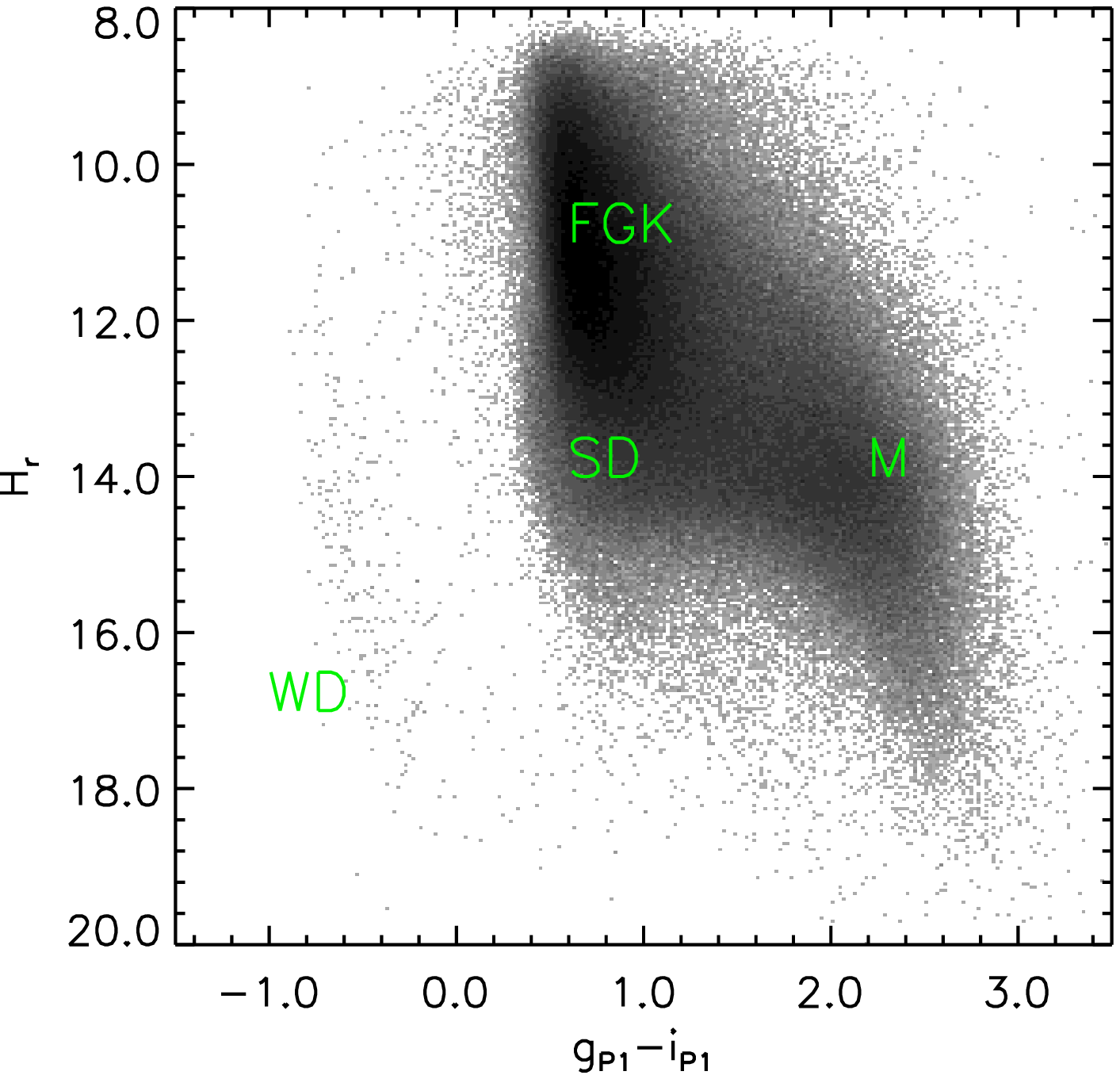}&\includegraphics[scale=0.5]{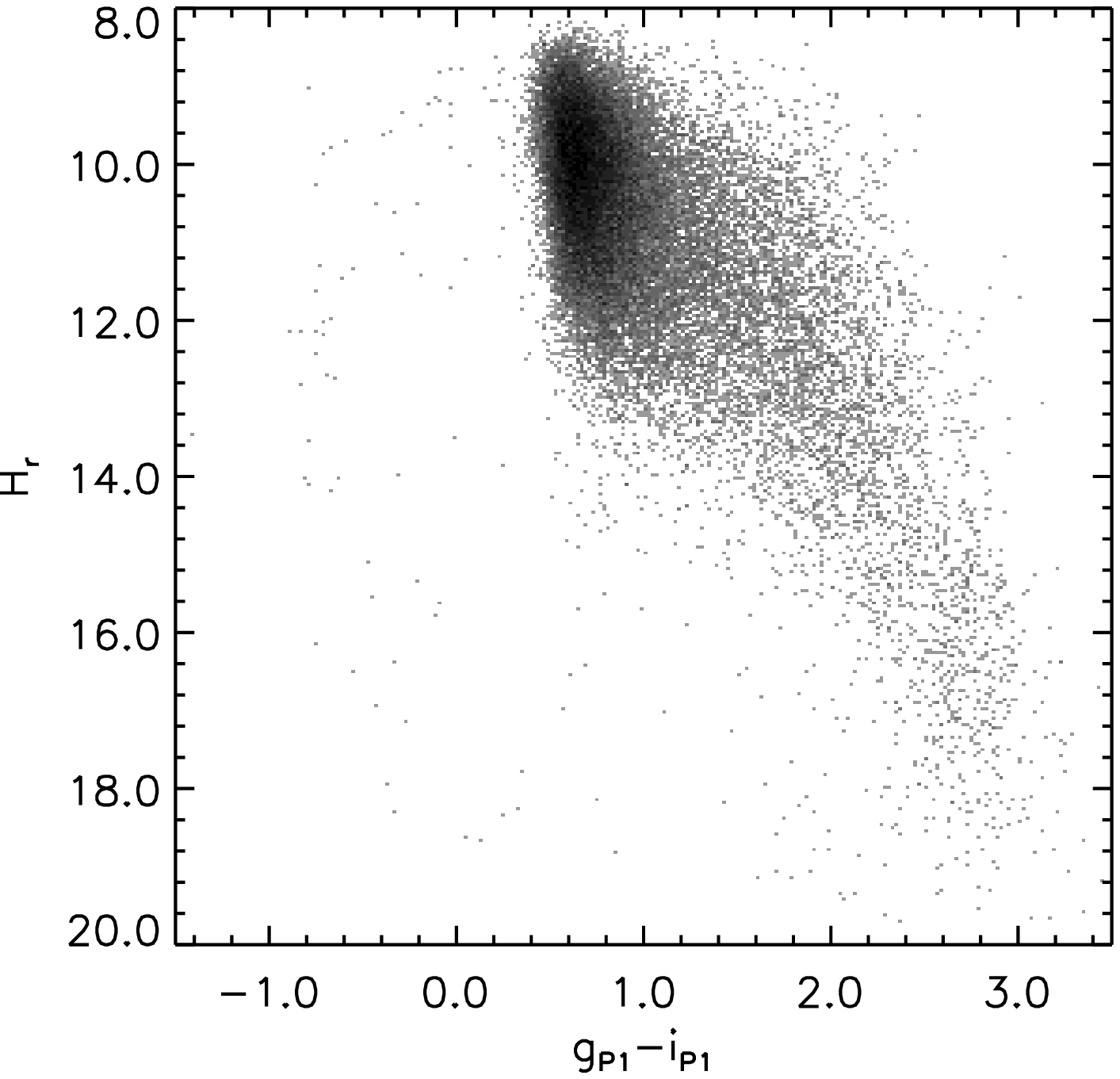}
\end{tabular}
\caption{\label{ps1_rpm} The $\log_{10}$ density distribution of reduced proper motion vs. colour for all sources with good quality 5$\sigma$ proper motions (left) and {\it Kepler} targets meeting the same criteria. Note the clear white dwarf (marked WD),  F, G, K dwarf (marked FGK), subdwarf (marked SD) and M dwarf (marked M) loci on the left-hand diagram. The right-hand plot is dominated by F, G and K stars, showing the known preference in {\it Kepler} target selection. }
\end{center}
\end{figure}

Proper motions are presented in Table~\ref{pm_tab}. In our remaining analysis we define significant proper motions as being more significant than 5$\sigma$ and good quality proper motions as being calculated from 10 or more measurements, having a USNO object within 6$\arcsec$ and one PS1 object within 3$\arcsec$,~ being fainter than $r_{P1}=14.5$\,mag. (i.e. not saturated), brighter than $r_{P1}=19$\,mag and having a reduced $\chi^2$ less than 4. This results in proper motions which are not confused by crowding, saturation or outliers and likely involving a time baseline of over 50 years. Figure~\ref{ps1_ccd} shows the distribution of all objects in our sample with unsaturated photometry (left) and photometry for unsaturated {\it Kepler} targets (right) in a $r_{P1}-i_{P1}$ vs.$g_{P1}-r_{P1}$ . This shows that there are relatively few {\it Kepler} targets which appear to be reddened beyond the stellar locus. Figure~\ref{ps1_rpm} shows reduced proper motion diagrams ($H_r = r_{P1}+5+5\log_{10}\mu$) for both all objects with good quality significant proper motions (left) and {\it Kepler} targets passing the same criteria (right). Clear F, G, K dwarf, M dwarf, white dwarf and subdwarf loci can be seen in the left-hand diagram, but the {\it Kepler} targets are disproportionally dominated by F, G, and K dwarfs. This is unsurprising as these were the primary target class in {\it Kepler}'s search for Earth-like planets \citep{Batalha2010}.

\begin{figure}
\begin{center}
\begin{tabular}{cc}

\includegraphics[scale=0.5]{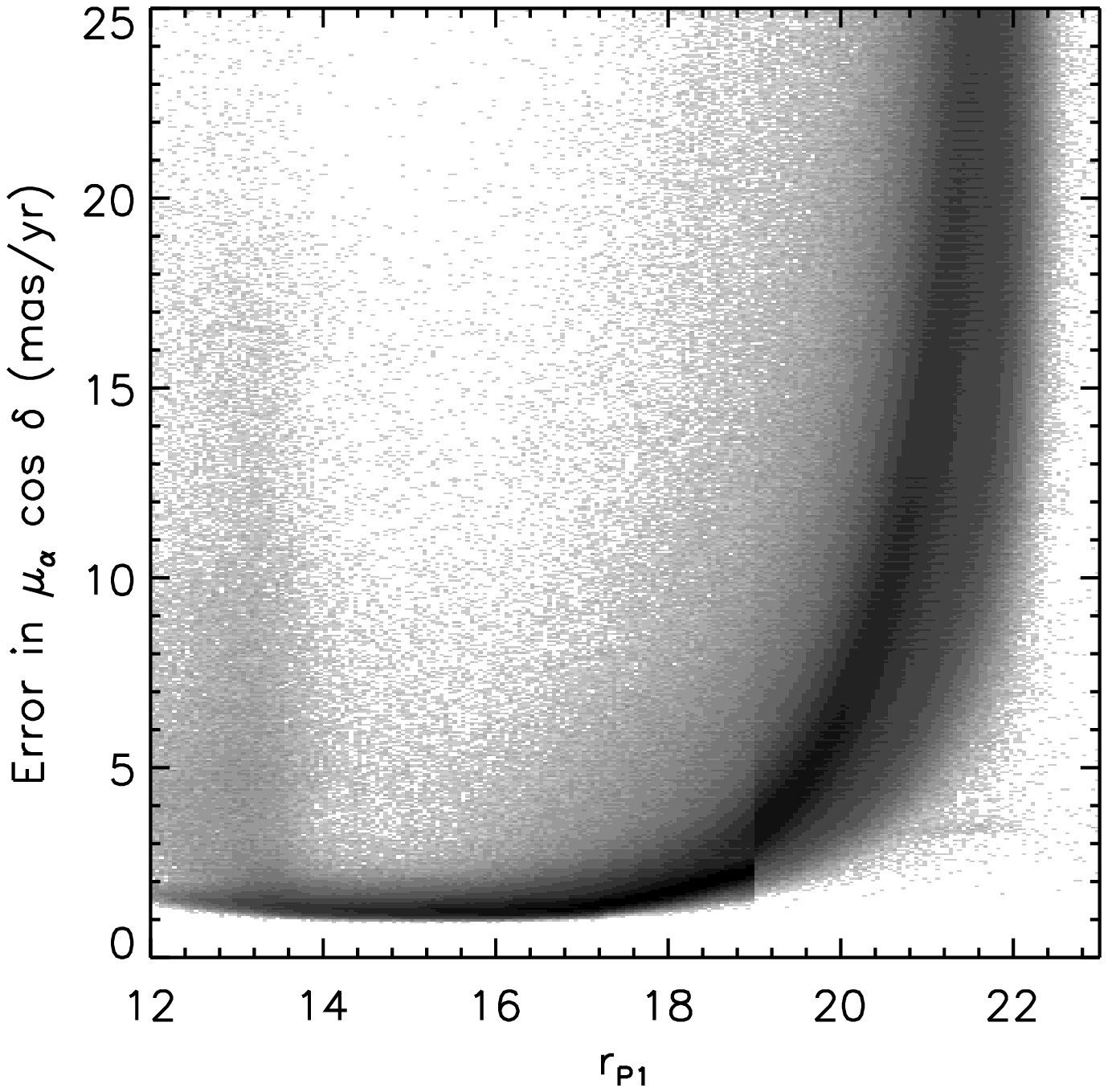}&
\includegraphics[scale=0.5]{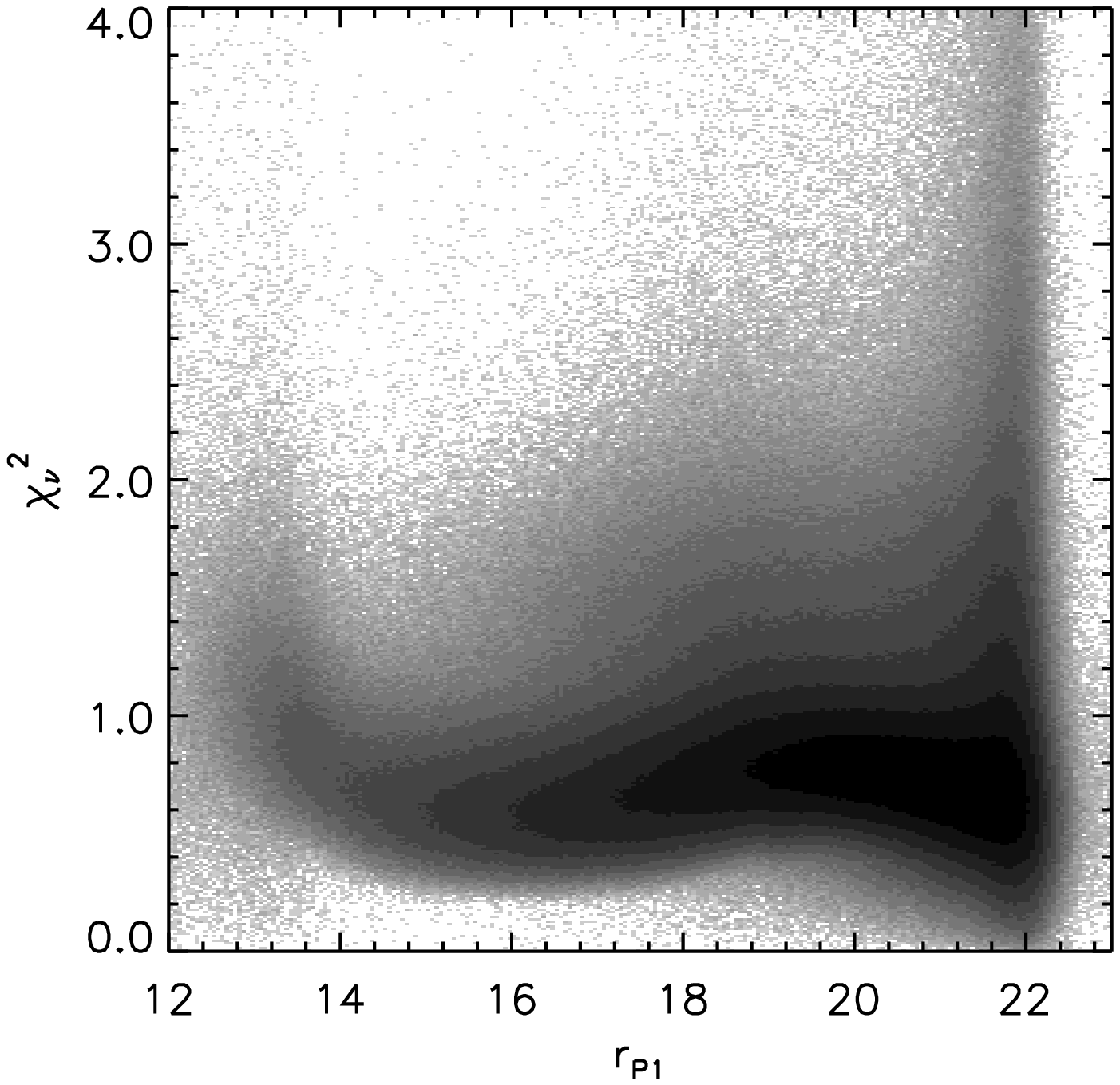}
\end{tabular}
\caption{\label{ps1_chi} The $\log_{10}$ density distribution of proper motion error vs. magnitude (left) and reduced $\chi^2$ vs. magnitude (right) for all sources. Note the discontinuity in proper motion error at $r_{P1}=19$. This is because we excluded matches with USNO detections at this magnitude to prevent matches with faint, spurious USNO detections. Note also both distributions become warped brighter than $r_{P1}=14$ due to saturation. Our reduced $\chi^2$ plot peaks at a value slightly below 1, indicating our astrometric errors are slightly over-estimated.}
\end{center}
\end{figure}

\clearpage

\begin{landscape}
\begin{table}
\caption{\label{pm_tab} Proper motions for unsaturated {\it Kepler} targets along with their Pan-STARRS\,1 magnitudes. The penultimate column flags if this object passes the "good" proper motion test previously set out in the text and the final column shows {\it Kepler} Objects of Interest which have passed or failed our vial proper motion inspection. The number of objects within 3 arcseconds and 6 arcseconds in both Pan-STARRS\,1 and USNO are also quoted. Objects marked with $\dag$ have multiple PS1 matches within 3$\arcsec$ but the astrometric solution selected is for an object within one arcsecond with the same magnitude as the Kepler target and is thus believed to be reliable. The full table is available electronically.}
\scriptsize
\begin{center}
\begin{tabular}{cccrrrcccccrrrrccc}
\hline
KID&R.A.&Dec.&$\mu_{ra}\cos\delta$&$\mu_{\delta}$&$n_{meas}$&$g_{P1}$&$r_{P1}$&$i_{P1}$&$z_{P1}$&$y_{P1}$&$\chi^2_{\nu}$&$n3_{PS1}$&$n6_{PS1}$&$n3_{USNO}$&$n6_{USNO}$&flag&visual\\
&\multicolumn{2}{c}{(Eq=J2000 Ep=2013.0)}&(mas/yr)&(mas/yr)&(mas/yr)&(mas/yr)&&(mag.)&(mag.)&(mag.)&(mag.)&(mag.)&\\
\hline
757450&19 24 33.03&+36 34 38.6&12.08$\pm$1.65&2.85$\pm$1.54&59&15.773$\pm$0.001&15.087$\pm$0.001&14.844$\pm$0.001&14.737$\pm$0.001&14.679$\pm$0.001&0.485&1&1&1&1&1&1\\
891916&19 23 49.98&+36 41 11.8&8.79$\pm$1.61&2.26$\pm$1.5&57&15.179$\pm$0.001&14.691$\pm$0.001&14.512$\pm$0.001&14.439$\pm$0.001&14.392$\pm$0.001&0.626&1&2&1&1&1&\\
892718&19 24 34.01&+36 38 53.9&$-$4.83$\pm$1.39&$-$4.04$\pm$1.27&65&16.515$\pm$0.002&15.842$\pm$0.001&15.586$\pm$0.001&15.451$\pm$0.001&15.373$\pm$0.002&0.553&1&1&1&2&0\\
892772&19 24 36.80&+36 40 43.8&$-$8.16$\pm$1.3&0.95$\pm$1.19&74&15.77$\pm$0.001&15.05$\pm$0.001&14.772$\pm$0.001&14.627$\pm$0.001&14.525$\pm$0.001&0.72&1&2&1&1&1&1\\
892832&19 24 39.15&+36 40 27.5&$-$13.11$\pm$1.32&$-$12.47$\pm$1.2&81&16.304$\pm$0.002&15.624$\pm$0.001&15.367$\pm$0.001&15.252$\pm$0.001&15.177$\pm$0.001&0.326&1&1&1&2&0\\
892834&19 24 39.20&+36 37 39.3&2.38$\pm$1.39&2.28$\pm$1.28&69&15.919$\pm$0.001&15.134$\pm$0.001&14.863$\pm$0.001&14.731$\pm$0.001&14.657$\pm$0.001&0.617&1&2&1&1&1\\
892882&19 24 41.58&+36 41 54.1&1.89$\pm$2.47&8.43$\pm$2.23&23&15.53$\pm$0.001&14.862$\pm$0.001&14.638$\pm$0.001&14.542$\pm$0.001&14.483$\pm$0.001&1.363&1&3&1&1&1\\
892911&19 24 43.13&+36 40 14.1&10.57$\pm$1.93&0.18$\pm$1.86&18&16.081$\pm$0.001&15.601$\pm$0.001&15.433$\pm$0.001&15.366$\pm$0.001&15.314$\pm$0.002&1.046&2&2&1&1&0\\
892946&19 24 45.03&+36 40 21.0&8.44$\pm$1.85&3.73$\pm$1.76&25&16.237$\pm$0.001&15.804$\pm$0.001&15.663$\pm$0.001&15.622$\pm$0.002&15.586$\pm$0.002&1.285&1&2&2&2&0\\
893033&19 24 49.82&+36 40 03.8&19.18$\pm$1.32&$-$8.03$\pm$1.21&77&15.976$\pm$0.001&15.197$\pm$0.001&14.899$\pm$0.002&14.776$\pm$0.001&14.699$\pm$0.001&0.442&1&1&1&2&0\\
\hline
\end{tabular}
\normalsize
\end{center}
\end{table}
\end{landscape}
\subsection{Verification of proper motions}
\begin{figure}
\begin{center}
\begin{tabular}{cc}
\includegraphics[scale=0.5]{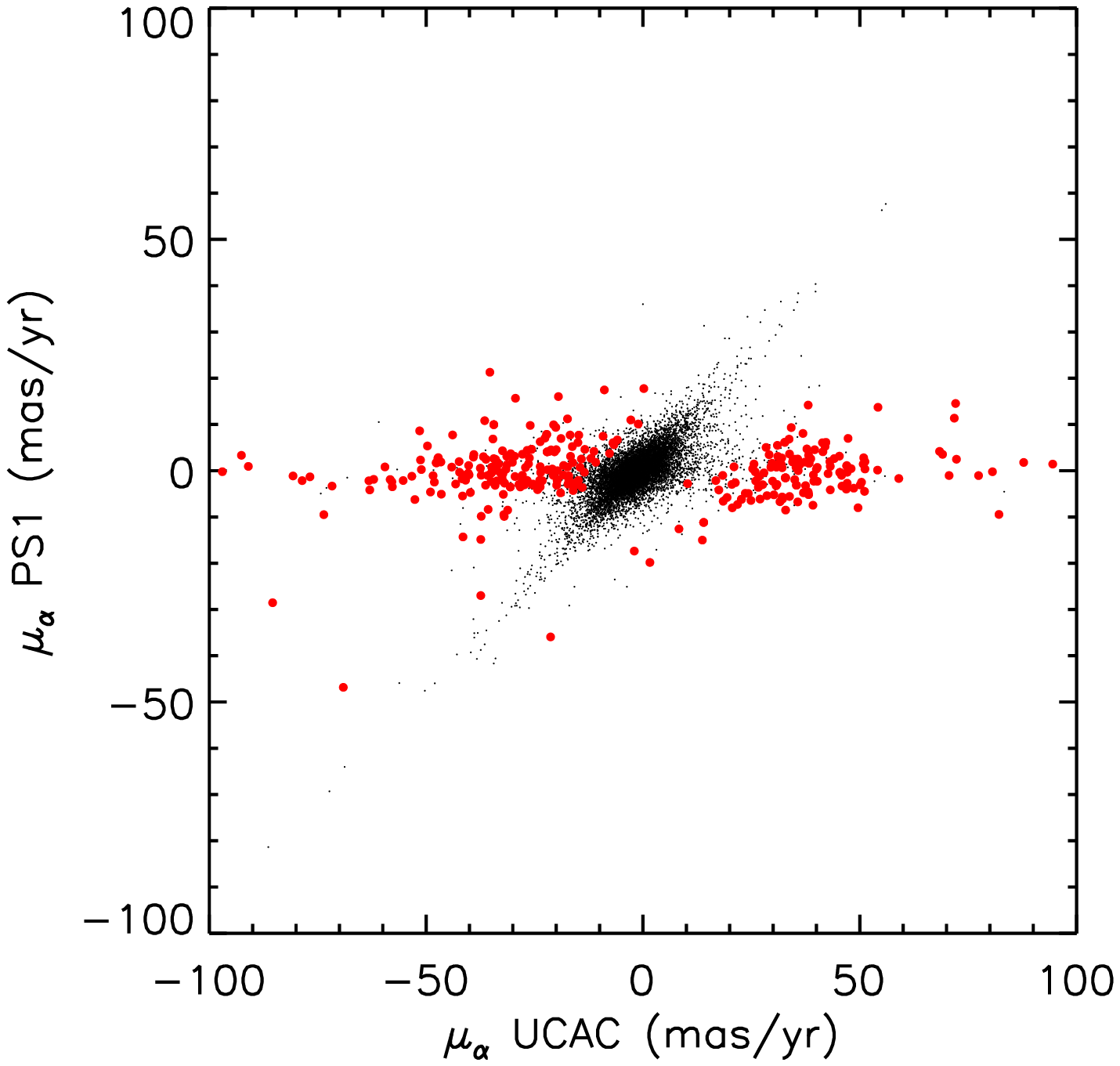}&\includegraphics[scale=0.5]{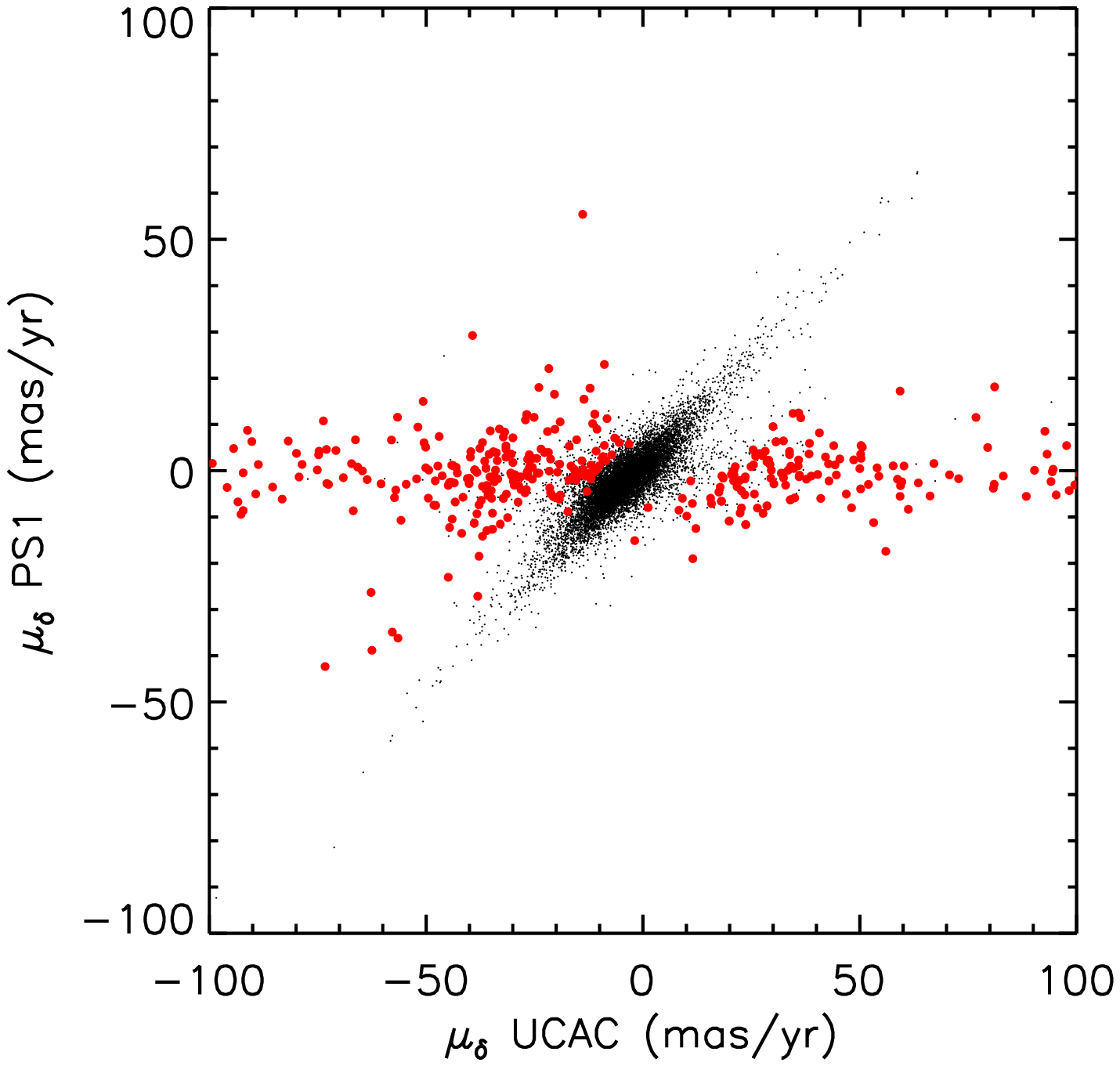}
\end{tabular}
\caption{\label{ps1_ucac} Comparison of our Pan-STARRS\,1 proper motions with those from the UCAC4 catalogue \protect\citep{Zacharias2013}. Only points which pass our proper motion quality cuts are included. Red points show 5$\sigma$ disagreements between our proper motions and UCAC4 in R.A (left panel) or Dec. (right panel). The majority of data points fall along a one-to-one relation between the two proper motions. A number of objects have low proper motions in our calculations but high UCAC proper motions}
\end{center}
\end{figure}

To verify our proper motions we compared our selection of reliable proper motions with the values from the UCAC4 survey (\citealt{Zacharias2013}; see Figure~\ref{ps1_ucac}). This dataset does not use Schmidt survey plates so it represents an independent dataset. We found that our values generally compared well with the UCAC values. However, there was a small number (3.3\%) of objects for which we measure much lower proper motions compared to UCAC. A visual inspection of a subset of these objects suggests that these objects do not have significant proper motions and often have another star at a 6--10$\arcsec$ distance suggesting that these high proper motions are an artefact of object confusion in UCAC.
As an additional check, we individually inspected the astrometric solutions for 3800 Kepler Objects of Interest listed in \cite{Rowe2015} for which we had a proper motion solution. We then made a by-eye judgement on how reliable the proper motion solution was compared to the distribution of astrometric points. We found that 94\% of our proper motions passed this test. Of the objects which failed visual inspection, approximately three quarters had proper motions which would not meet our reliability cuts (high reduced $\chi^2$, few measurements, etc.). This KOI visible checking is listed as a parameter in Table~\ref{pm_tab}.

\subsubsection{White dwarf {\it Kepler} Targets}
Proper motion catalogues with high levels of contamination due to spurious proper motions often present reduced proper motion diagrams with white dwarf loci which are heavily contaminated by FGK stars with erroneously high proper motion measurements. Hence the population of the white dwarf locus is a useful test of the reliability of a proper motion catalogue. To identify a sample of {\it Kepler} target white dwarfs we selected unsaturated objects which had good, significant proper motions, were redder than $g_{P1}-i_{P1}=-2$ (to exclude objects with erroneously blue photometry), met the condition $H_r>10\times(g_{P1}-i_{P1})+12$ and whose proper motion solutions passes a visual inspection. These objects are shown in Table~\ref{WD_cand}. Most of the intrinsically brighter objects passing this cut (those with low reduced proper motion) are sdO or sdB stars based on literature classifications (see Table~\ref{WD_cand}). This cut also misses intrinsically bright white dwarfs which will be distant enough that they would not pass our proper motion significance cut. Twelve of our fifty candidate white dwarfs have been previously identified as white dwarfs, dwarf novae or cataclysmic variables. We do not see significant contamination from objects spectrally classified as FGK stars.
\begin{table}
\footnotesize
\caption{\label{WD_cand} $^1$\protect\cite{Ostensen2010}, $^2$\protect\cite{Ostensen2011a}, $^3$\protect\cite{Reed2011}, $^4$\protect\cite{Girven2011}, $^5$\protect\cite{McNamara2012}, $^6$\protect\cite{Hoffmeister1966}, $^7$\protect\cite{Ostensen2011}, $^8$\protect\cite{Howell2013}, $^9$\protect\cite{Bischoff-Kim2011}, $^{10}$\protect\cite{Scaringi2013a}, $^{11}$\protect\cite{Feldmeier2011}, $^{12}$\protect\cite{Baran2011}, $^{13}$\protect\cite{Gianninas2011},$^{14}$\protect\cite{Hog2000}, $^{14}$\protect\cite{Greiss2014}.}
\begin{tabular}{lllr}
\hline
{\bf {\it Kepler}\_ID} &{\bf Other name}&{\bf Literature}&{\bf Reference}\\
&&{\bf classification}&\\
\hline

  3353239 &2MASS J19364633+3825268 &sdB& $^2$\\
  3527751 &2MASS J19033701+3836126 &sdB& $^3$\\
  3561157&  & \\
  3629120 &SDSS J190410.70+384518.1 &WD cand& $^4$\\
  3663415&  & \\
  3751235 &  & \\
  4357037 &SDSS J191719.16+392718.8 & WD cand& $^4$\\
  5077438&TYC 3123-1545-1&&$^{14}$\\
  5544772 &  &  &$^5$\\
  5951261 &  & &$^5$\\
  6042560 &  & \\
  6212123 &  & \\
  6672883 &  & \\
  7346018 &  & \\
  7594781 &  & \\
  7659570 & V* V344 Lyr & DN&  $^6$\\
  7797992 &  & \\
  8075923 &  & \\
  8077281 & 2MASS J18501686+4358284 & B&$^1$\\
  8210423 &  & \\
  8244398 &  & \\
  8395780 &  & \\
  8420780 &  GALEX J192904.6+444708 & DB &$^7$\\
  8490027 &  & CV cand& $^8$ \\
  8626021 & WD J1929+4447 &DB&  $^9$\\
  9071514 &&CV& $^{10}$ \\
  9139775 &  & DA $^1$\\
  9228724 &  & \\
  9391127 &  & \\
  9535405 &  &  DA&  $^{11}$\\
  9569458 &  & sdB& $^1$\\
  9818160 &  & \\
  10066680&&&\\
  10081214 &  & \\
  10149875 &  & \\
  10290697 &  & \\
  10462707 &  &  sdB&$^2$\\
  10590313 &  &LSPM J1910+4750  & \\
  11176123 &  & \\
  11191648 &  & \\
  11351218 &  &&$^5$ \\
  11357853 &  &sdOB& $^1$ \\
  11402999 &  & \\
  11509531 &  & \\
  11558725 & 2MASS J19263411+4930296&sdB & $^{12}$ \\
  11604781 &  2MASS J19140898+4936410 &DA& $^2$ \\
  11822535 & WD 1942+499& DA1.4 & $^{13}$\\
  11911480 &  &  DA/ZZ Ceti&$^{14}$\\
  12021724 &2MASS J19441275+5029393&sdB & $^2$ \\
  12353867 &  & \\
  \hline
\end{tabular}
\normalsize
\end{table}

\section{Identification of likely binary companions}
\subsection{Probabilistic binary selection}
Stars in wide binary systems are typically found due to their common proper motion. Identification of such pairings is complicated by the presence of nearby stars, which by chance, have the same proper motion as each other despite having no physical connection. These coincident pairings are often removed with hard cuts or using by-eye judgements. Two better defined processes for binary star selection come from the work of \cite{Lepine2007} and \cite{Dhital2010}. The former empirically defines a region of proper motion and separation space where coincident pairings are unlikely. The latter method uses a model of the Galaxy to define the probability that an unrelated star would have a particular proper motion and position difference from the target primary star. Here we define a hybrid of these two systems which calculates the probability that an object is a true binary based on a estimate of the background star population, but unlike Dhital (2010), our coincident pairing estimates are purely empirical. However unlike the two previously mentioned methods, we take the likely distribution of the wide binary population into account as well. Similar methods of separating objects from a background population have been used to identify members of open clusters using their proper motion such as \cite{Deacon2004}, \cite{Kraus2007} and \cite{Sanders1971}. While these methods fit distribution functions to the background population, we use the \cite{Lepine2007} method of offsetting the position of the target primary and then calculating the density of objects with similar proper motions, positions, distances and masses to the candidate secondary.  Appendix~\ref{probmaths} outlines the mathematics used in our probability calculations.
\subsection{Testing the binary probability algorithm}
\label{bin_test}
To test that our method for binary association probability is accurate we used a sample of bright stars from our {\it Kepler} proper motion catalogue. These had proper motions from UCAC4 \citep{Zacharias2013} or PPMXL \citep{Roeser2010} and photometric distances and mass estimates calculated using the methods of \cite{Kraus2007} and \cite{Kraus2014a}. We assumed a log-flat separation distribution \citep{Opik1924} \footnote{ This is flatter than the typical log-normal separation distribution used for binaries closer than 100\,AU but is appropriate as the binaries we identify in this work are typical a few thousand AU in separation}, a flat mass ratio distribution \citep{Raghavan2010} and set the binary fraction wider than 100\,AU to be 25\%. This latter number is similar to the number of binaries wider than this limit found around nearby solar-type stars by \cite{Raghavan2010}. We did not set a lower proper motion limit and considered only pairs with observed separations below five arcminutes and projected separations below 10,000\,AU. We used this technique to calculate companionship probabilities for these pairs and then flagged a subsample of these pairs for follow-up observations.

\begin{figure}
\begin{center}
\includegraphics[scale=0.5]{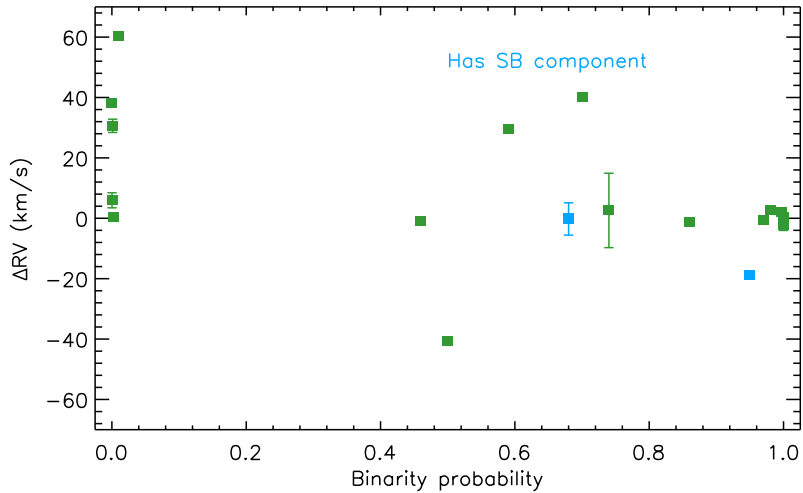}
\caption{\label{p_bin_rv} The observed radial velocity differences for a series of candidate binaries with a range of calculated binary probabilities. These objects had proper motions from UCAC4 and SED fits from the catalogue built for \protect\cite{Kraus2014a}. Two of our pairs have a component which is a spectroscopic binary. This will induce offsets in their radial velocity. Note also the large fraction of objects which have high binarity probabilities and common radial velocities.}
\end{center}
\end{figure}

A total of 26 wide binary candidate systems (52 stars) were observed with the Tull Coude spectrograph \citep{Tull1995} at the Harlan J. Smith 2.7m Telescope at McDonald Observatory on the nights of August 8, 9, 10, and 11, 2014. Exposure times varied from 120s to 1200s, which was sufficient to achieve a median SNR (per resolving element) of $>30$. The resulting spectra cover 3800\AA\ to 10500\AA, with a gap from 4550\AA\ to 4750\AA\ due to internal reflection, and at a resolution of $R\simeq60,000$. Pair components were always observed within minutes of each other to reduce systematic sources of error (e.g., atmospheric changes, telescope flexure, EarthÕs motion). 
Data were reduced using standard IRAF reduction tools, including bias and flat field correction, extraction of the 1 dimensional spectrum, and wavelength calibration. An additional wavelength correction was applied by cross correlating the observed telluric lines to a model atmosphere as described in \citet{Gullikson2014}. We cross-correlated the observations of our target pair components with each other to determine the radial velocity difference. These values for our observed pairs (along with their binarity probability $p_{bin}$) are listed in Table~\ref{RV_obs}. The errors in radial velocity difference come from the cross correlation (with a 4$\sigma$ clipping) but do not include other potential sources of random and systematic error (e.g, additional undetected companions) and hence they are likely to be an underestimate of the error of the radial velocity difference. Figure~\ref{p_bin_rv} shows our results with two pairs flagged as containing a spectroscopic binary (i.e. are hierarchical triples). These were KIC~9085834 and KIC~11017620, which both produced a double-peaked correlation function or showed pairs of stellar lines. Nine of the pairs we classify as having $p_{bin} > 0.6$ have radial velocity differences less than 4\,km/s. One of these contains spectroscopic binary while another three objects in the same probability range have radial velocities which are discrepant (excluding spectroscopic binaries). Of the five low probability objects we observed all but one have discrepant radial velocities. This suggests that our method can reliably select true co-moving systems.
\begin{table}
\scriptsize
\caption{\label{RV_obs} Wide binary candidates with different binary probabilities $p_{bin}$ which had their RV differences measured.  Key: $^*$Spectroscopic binary (SB2), \protect\dag Rapid rotator, Proper motions from $^1$\protect\cite{Zacharias2013}, $^2$\protect\cite{Roeser2010}.}
\begin{tabular}{llcccccc}
\hline
KIC&
$\mu_{\alpha} \cos \delta$&
$\mu_{\delta}$&
SpT&
Distance modulus&
$r$&
$\Delta$RV&
$p_{bin}$\\
&
(mas/yr)&
(mas/yr)&
(subtypes)&
(mag)&
(arcsec)&
(km/s)\\
\hline
2696938&5.2$\pm$2.0$^2$&$-$18.3$\pm$2.0&F6.7$\pm$1.5&6.31$\pm$0.22&5.4&$-$0.55$\pm$0.85&1.00\\
2696944&1.0$\pm$1.8$^1$&$-$20.6$\pm$2.4&F4.9$\pm$1.4&6.53$\pm$0.22\\
\hline
2992956&16.5$\pm$1.4$^2$&50.2$\pm$1.4&G6.5$\pm$1.45&5.13$\pm$0.2&16.9&0.47$\pm$0.07&1.00\\
2992960&14.5$\pm$1.4$^2$&51.8$\pm$1.4&G6.9$\pm$1.45&5.3$\pm$0.19\\
\hline
4243796&$-$1.2$\pm$0.7$^1$&$-$18.7$\pm$1.1&K4.1$\pm$0.6&3.91$\pm$0.07&154.3&0.52$\pm$0.05&0.00\\
4346953&$-$9.5$\pm$2.0$^1$&$-$17.7$\pm$2.5&K7.2$\pm$0.15&4.19$\pm$0.07\\
\hline
5790787&4.5$\pm$0.9$^1$&10.6$\pm$0.8&F6.9$\pm$1.45&7.45$\pm$0.2&27.6&40.07$\pm$1.21&0.70\\
5790807&5.1$\pm$0.6$^1$&9.9$\pm$1.0&F3.6$\pm$1.35&6.52$\pm$0.23\\
\hline
6934317&$-$6.3$\pm$0.8$^1$&$-$15.5$\pm$0.9&G9.9$\pm$1.2&6.46$\pm$0.12&44.7&38.29$\pm$0.11&0.00\\
7019341&$-$11.7$\pm$0.6$^1$&$-$13.1$\pm$0.6&G2.5$\pm$0.75&6.66$\pm$0.05\\
\hline
7090649&16.7$\pm$1.8$^2$&13.6$\pm$1.8&G7.0$\pm$2.6&5.97$\pm$0.28&9.3&$-$2.11$\pm$1.91&1.00\\
7090654&17.5$\pm$1.2$^1$&13.3$\pm$1.2&F7.2$\pm$1.85&5.80$\pm$0.23\\
\hline
7748234&$-$6.2$\pm$1.6$^2$&3.9$\pm$1.6&F8.2$\pm$0.2&7.14$\pm$0.03&38.0&2.6$\pm$12.32&0.74\\
7748238&$-$7.6$\pm$0.9$^1$&0.7$\pm$0.7&F0.0$\pm$1.65&6.8$\pm$0.17\\
\hline
8123664&$-$5.0$\pm$1.5$^2$&$-$6.4$\pm$1.5&F1.7$\pm$5.15&5.96$\pm$2.5&5.6&8.63$\pm$0.62&0.00\\
8123668\dag&$-$6.6$\pm$1.5$^2$&$-$16.9$\pm$1.5&F7.5$\pm$4.75&7.70$\pm$0.49\\
\hline
8619322&$-$4.7$\pm$0.9$^1$&$-$4.8$\pm$1.0&K1.8$\pm$0.85&3.62$\pm$0.11&110.8&$-$40.63$\pm$0.07&0.51\\
8683779&$-$2.5$\pm$1.3$^1$&$-$6.3$\pm$0.9&K4.4$\pm$0.6&3.87$\pm$0.06\\
\hline
9085833\dag&$-$3.8$\pm$0.6$^1$&$-$15.9$\pm$0.8&F4.8$\pm$1.35&7.53$\pm$0.21&20.1&$-$0.22$\pm$5.35&0.68\\
9085834$^*$&$-$4.3$\pm$1.6$^1$&$-$13.3$\pm$3.4&G2.3$\pm$3.6&8.36$\pm$0.32\\
\hline
9579191&$-$13.1$\pm$2.1$^1$&10.2$\pm$2.0&G5.0$\pm$1.55&7.05$\pm$0.15&20.3&$-$1.27$\pm$0.32&0.0.86\\
9579208&$-$11.2$\pm$0.8$^1$&8.8$\pm$1.0&F6.4$\pm$1.45&6.39$\pm$0.21\\
\hline
9595822&$-$2.3$\pm$1.0$^1$&$-$5.3$\pm$0.9&K3.5$\pm$0.25&4.67$\pm$0.03&69.0&29.64$\pm$0.11&0.59\\
9534041&$-$3.5$\pm$1.1$^1$&$-$6.9$\pm$1.0&K1.3$\pm$0.9&4.92$\pm$0.12\\
\hline
9655101&$-$4.4$\pm$0.6$^1$&$-$9.3$\pm$0.9&K2.2$\pm$0.65&4.73$\pm$0.08&65.0&2.7$\pm$0.1&0.98\\
9655167&$-$4.7$\pm$0.7$^1$&$-$8.7$\pm$1.1&K3.0$\pm$0.45&4.68$\pm$0.05\\
\hline
9777293&$-$9.8$\pm$0.5$^1$&11.5$\pm$0.7&K1.6$\pm$0.85&4.77$\pm$0.11&48.9&2.13$\pm$0.08&1.00\\
9777355&$-$9.4$\pm$1.3$^1$&11.6$\pm$0.8&K2.0$\pm$0.7&4.79$\pm$0.09\\
\hline
9912680&11.6$\pm$2.0$^1$&1.0$\pm$0.6&G0.4$\pm$2.7&6.53$\pm$0.21&7.1&$-$0.5$\pm$0.27&0.97\\
9912690&14.5$\pm$1.3$^1$&2.5$\pm$1.2&G0.9$\pm$4.6&7.39$\pm$0.4\\
\hline
10669568&$-$0.1$\pm$2.0$^1$&$-$10.9$\pm$3.4&K4.2$\pm$0.6&5.68$\pm$0.06&30.1&60.48$\pm$0.17&0.01\\
10669590&$-$10.6$\pm$3.5$^1$&$-$11.4$\pm$2.3&K4.5$\pm$0.35&5.91$\pm$0.04\\
\hline
11017620$^*$&$-$9.4$\pm$2.1$^1$&$-$19.0$\pm$0.5&F7.1$\pm$1.95&6.53$\pm$0.24&8.9&$-$18.93$\pm$0.6&0.95\\
11017626&$-$4.8$\pm$1.2$^1$&$-$17.5$\pm$0.9&F8.8$\pm$4.2&7.21$\pm$0.38\\
\hline
11551404&17.6$\pm$1.5$^1$&12.6$\pm$1.1&K2.6$\pm$0.6&4.66$\pm$0.08&55.7&30.61$\pm$2.2&0.00\\
11551430&11.2$\pm$1.7$^1$&20.6$\pm$1.4&K0.5$\pm$1.05&4.88$\pm$0.13\\
\hline
12253474&$-$11.4$\pm$2.3$^1$&$-$8.1$\pm$0.5&G0.7$\pm$4.65&7.7$\pm$0.4&18.0&$-$1.02$\pm$0.17&0.45\\
12253481&$-$5.4$\pm$2.2$^1$&$-$7.0$\pm$1.0&A6.6$\pm$5.05&8.23$\pm$0.61\\
\end{tabular}
\normalsize
\end{table}

\subsection{Application to Pan-STARRS\,1 {\it Kepler} targets}
\begin{figure}
\begin{center}
\includegraphics[scale=0.7]{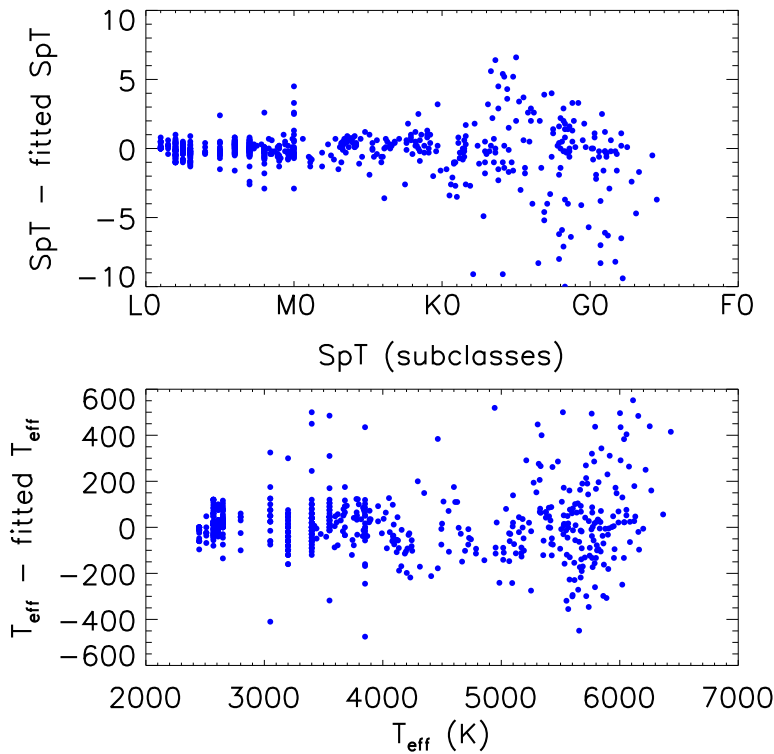}
\caption{\label{sed_comp} A comparison between the measured spectral types and effective temperatures for our calibration stars and those derived from fitting. Note the objects shown here have at least four photometric measurement and exclude out M9.5 and L0 calibration objects as we do not have an L0 template to fit these against. In the upper plot a positive y-axis value indicates that our spectral type estimate from SED fitting is earlier than that derived from spectroscopy. Many of the M dwarfs have effective temperatures derived from their spectral types, hence the discrete temperatures in this region of the plot}
\end{center}
\end{figure}

The testing of our binary selection algorithm outlined in Section~\ref{bin_test} only applied to objects with astrometry in UCAC\,4 and used only photographic plate and 2MASS photometry. We can widen our binary search by looking at fainter companions which have reliable Pan-STARRS\,1 photometry. To do this we need a series of Spectral Energy Distribution (SED) templates using the process outlined in Appendix~\ref{sed_sect}. A by-product of this search allowed us to estimate accurate transformations from Pan-STARRS\,1 colours to $K_p$ magnitudes. This process is outlined in Appendix~\ref{kp_sect}. We performed our SED fits using the same process as described by \cite{Kraus2007} fitting a reduced $\chi^2$ for all good data points, removing the most discrepant data point around the best fit (the fit with the lowest reduced $\chi^2$) if it is more discrepant than 3$\sigma$, refitting, redoing our clipping of the most discrepant point with the same conditions as before and finally refitting. We do not clip if it would leave fewer than four good data points and we not do an initial fit if we have fewer than this number. Note that for the Pan-STARRS\,1 photometry we exclude filters where the object is marked as extended and has fewer than four detections. We also exclude data from filters above the following saturation magnitude limits: $g_{P1}=14.5$, $r_{P1}=14.5$, $i_{P1}=14.5$, $z_{P1}=14.5$, $y_{P1}=14.5$ and $J_{MKO}=10.5$. This last saturation limit is the same as that used in \cite{Deacon2009a}. It should be stated that our SED templates (like those used in \citealt{Kraus2014a}) assume the target is a dwarf. Hence this method may miss or incorrectly match binaries containing one or more evolved stars. To test the accuracy of our SED fitting method, we ran our calibration sample SEDs (as defined in Appendix~\ref{sed_sect}) through our fitting procedure. The results of these fits are shown in Figure~\ref{sed_comp}. This plot only includes astronomical objects with actual Pan-STARRS\,1 measurement used for calibration not the synthetic Pan-STARRS\,1 magnitudes used to model hot stars. 

The method produces reliable results for a range of spectral types, but with larger scatter for F and G type stars than K and M. Using a robust estimate of the scatter based on median absolute deviation we estimate that the standard deviation for our method is 2.5 subclasses for F and G stars and 0.6 subclasses for K and M stars. This equates to a scatter of 190\,K and 75\,K respectively. \cite{Kraus2014a} note that their SED fitting is accurate to less than one subclass. There is also the fact that the underlying calibration objects have measurement errors on their effective temperatures and/or those temperatures are derived from spectral types which are typically quoted in discrete steps of half a subclass. For example our F and G calibration stars which are mostly drawn from \cite{Huber2014} have typical errors on their measured effective temperatures of about 110\,K equating to 1.5 subclasses. This means that our quoted scatter in our measurements is likely to be around 180\,K or 2.0 subclasses for F and G stars. We choose not to alter our scatter on K and M stars as we do not have a formal error on spectral type (and hence effective temperature) for many of these calibration objects. Our SED fitting procedure produces error estimates based on the calculated reduced $\chi^2$ distribution. However these are often extremely small confidence ranges compared to the scatter we have derived here. Hence when an object has a confidence range in spectral type that falls below our calculated empirical scatter, we substitute our empirical scatter measurements to estimate one sigma confidence regions for our calculated distance moduli and bolometric magnitude. This results in a typical uncertainty in bolometric magnitude of 0.4\,mag.

We included astrometry and SED fits for stars brighter than our Pan-STARRS\,1 saturation limit of $r_{P1}<14.5$ by including astrometry from either the UCAC4 catalogue \citep{Zacharias2013} or the PPMXL catalogue \citep{Roeser2010}. Where stars had an entry in both catalogues we selected the data source with the lowest quoted error on proper motion. To avoid losing objects around our saturation boundary we did not set any limits on magnitude, we simply excluded UCAC4 or PPMXL measurements of any object with an unsaturated Pan-STARRS\,1 proper motion measurement in our catalogue. We then added SED fits from \cite{Kraus2014a} giving us proper motion, spectral type and distance information for most bright stars in the {\it Kepler} field.

\begin{figure}
\begin{center}
\includegraphics[scale=0.7]{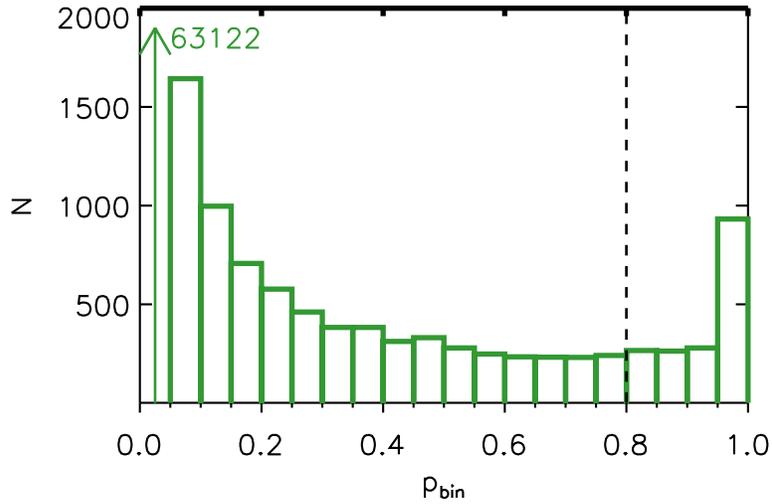}
\caption{\label{bin_prob} The distribution of binary probabilities for wide binaries with at least one {\it Kepler} target star. The arrow indicates the very large number of low probability pairings. The dashed line shows our 80\% threshold for high probability binaries}
\end{center}
\end{figure}

We used our combined catalogue as a basis for our wide binary selection. From it we selected possible companions to {\it Kepler} targets separated by less than five arcminutes from their primary and with distance moduli which differed by less than two magnitudes. We then ran our binary probability algorithm using our quoted errors on distance modulus and proper motion for all our possible pairings. A histogram of our membership probabilities is shown in Figure~\ref{bin_prob}. We have a large number of low probability pairings and an excess of high probability pairings with $p_{bin}>0.8$.

\subsubsection{Exclusion of giants}
One underlying assumption in our SED fits is that all the stars for which we derive distance moduli are dwarfs. This is clearly untrue as many stars in the {\it Kepler} field will be giants. \cite{Mann2012} showed that bright, red targets in the {\it Kepler} field are typically giants. To remove giants from our binary sample we set two cuts, one using $g$, $r$ and $D_{51}$ photometry from the {\it Kepler} Input Catalogue \citep{Brown2011} and one using 2MASS photometry. Both these use the separation \cite{Mann2012} showed between the late-type giant and dwarf populations. We define giants as either having $J-K_s>1$ or having $g-D_{51}>0.25$ and $g-r<1.3$. The first cut used here would exclude very late-type companion to a {\it Kepler} target so we do not impose this 2MASS related cut on our secondary stars. As an additional check we looked at the {\it Hipparcos} parallaxes \citep{vanLeeuwen2007} of stars in our binary sample. Of the 9 stars with measured {\it Hipparcos} parallaxes, none had distances which were significantly discrepant from our photometric distance estimates. 

We have endeavoured to minimise giant contamination but as the distance moduli for these binaries use the assumption that they are dwarfs, anyone using individual binaries from this sample should be aware that giant contamination may not have been completely eradicated and some giants may remain in the sample. We urge particular caution for the systems with primaries KIC 6367993 and KIC 10592818 as these have distance moduli less than one.

\section{The wide binary population in the {\it Kepler} field}
\begin{figure}
\begin{center}
\includegraphics[scale=0.5]{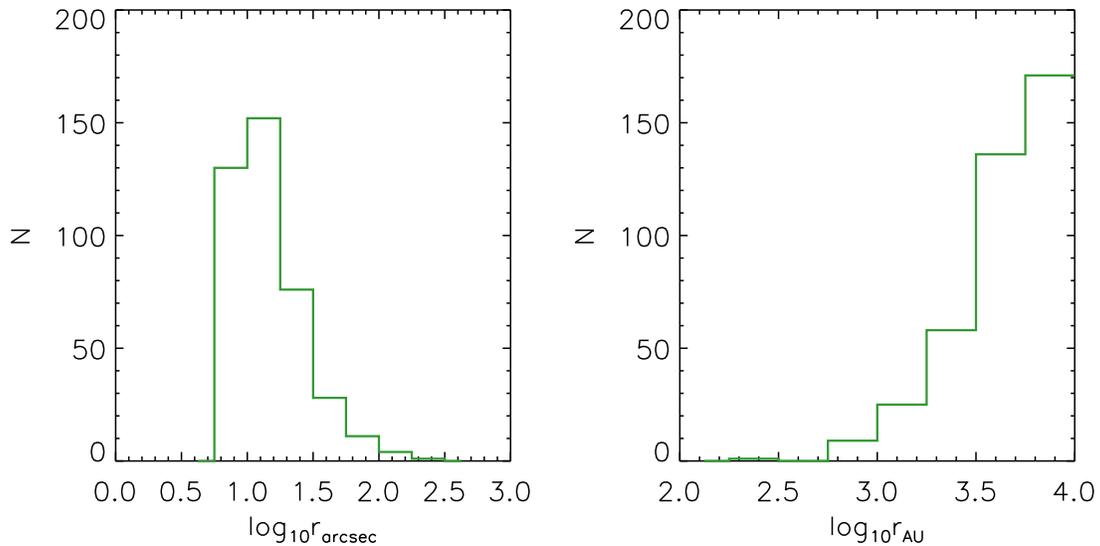}
\caption{\label{bin_sep} The properties of our binaries wider than 6$\arcsec$ with binary probabilities greater than 0.8. The separations are shown in arcseconds (left) and AU at the calculated distance of the {\it Kepler} target (right).}
\end{center}
\end{figure}

After excluding giant stars and removing binaries closer than 6$\arcsec$\footnote{This cut was to remove binaries where the presence of a close companion star would cause errors in the determination of exoplanet parameters and rotation periods.} we were left with Kepler targets in 401 binary systems which have binary probabilities above 80\%. We list these is Table~\ref{bin_tab} and show their projected physical and angular separations in Figure~\ref{bin_sep}. 
\subsection{The ages of wide binary components}
\begin{figure}
\begin{center}
\includegraphics[scale=0.5]{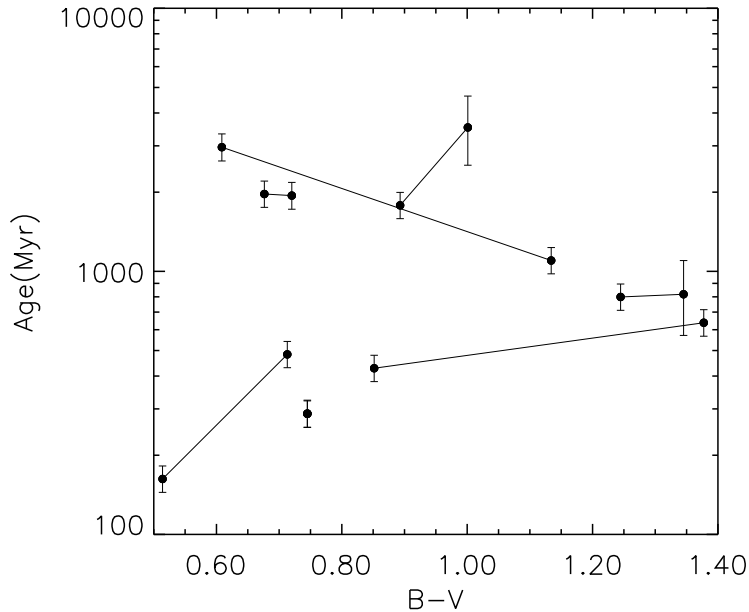}
\caption{\label{rot_plot} Ages calculated from gyrochronology for binaries with two periodic {\it Kepler} targets as components. The lines join the two components of each pair with horizontal lines representing good agreement between the ages of both components. The error bars on the ages include the 0.05\,dex errors suggested by \protect\cite{Mamajek2008}. What appears to be a single data point is the binary pair KIC 10913758 and KIC 10913762 which our SED fitting classifies as having the same spectral type and which have near-identical rotation periods.}
\end{center}
\end{figure}

Table~\ref{bin_tab} also contains rotation period estimates from \cite{McQuillan2014} for stars where this was measured. For pairs with two {\it Kepler} Targets with measured rotation periods and separations greater than 6$\arcsec$ we estimated the ages of both components using the age-rotation relations of \cite{Mamajek2008}. Rather than using measured $B-V$ values (which do not exist for most {\it Kepler} targets) for each target we converted our measured spectral types from SED fitting to colours using the field dwarf relation derived by \cite{Pecaut2013}. Figure~\ref{rot_plot} shows the ages derived for our pairs. Most have reasonable agreement between their ages. Four of our 7 binaries agree within one $\sigma$ (roughly as expected) with one further binary having ages within two $\sigma$. We would expect from our $>$80\% probability binary selection criteria that one or two of our seven proposed binaries would be not true physical pairs. Additionally, \cite{Mamajek2008} find that four of the solar-type binaries in their test sample of 17 systems have ages which are discrepant by more than 0.3\,dex. Hence our gyrochronology results do not indicate a larger than expected contamination in our binary pairs sample.

\clearpage
\begin{landscape}
\begin{table}
\scriptsize
\caption{\label{bin_tab} Our candidate binary systems, the full table will be available electronically. Rotations periods from \protect\cite{McQuillan2014} Key: $^a$ proper motion from \protect\cite{Zacharias2013}, $^b$SED fit calculated using the methods of \protect\cite{Kraus2014},$^c$ proper motion/SED fit from this work, $^d$ proper motion from \protect\cite{Roeser2010} .}
\begin{tabular}{rrccrrrrccrrrl}
\hline
KIC&KOI&
R.A.&Dec.&
$\mu_{\alpha} \cos \delta$&
$\mu_{\delta}$&
Distance modulus&
SpT&
$P_{rot}$&
\multicolumn{2}{c}{$r$}&
$p_{bin}$&WDS\\
&&\multicolumn{2}{c}{(Ep=2013.0, Eq=J2000)}&
(mas/yr)&
(mas/yr)&
(mag)&
(subtypes)&
(days)&
(arcsec)&
(AU)&Designation\\
\hline
1161145&&19:23:59.11&+36:52:31.2&$-$5.8$\pm$1.4$^a$&$-$3.5$\pm$1.2&7.26$\pm$0.07&G6.7$^b$&&10.7&3021&0.81&WDS 19240+3653B\\
1161137&&19:23:58.32&+36:52:26.2&$-$0.1$\pm$1.3$^a$&$-$0.4$\pm$1.6&7.22$\pm$0.07&G7.1$^b$&&&&WDS 19240+3653A\\
\hline
1295737&&19:26:30.04&+36:54:47.7&5.2$\pm$1.0$^a$&$-$5.8$\pm$2.2&8.36$\pm$0.21&F4.0$^b$&&13.1&6166&0.84\\
&&19:26:30.99&+36:54:41.2&0.7$\pm$1.8$^c$&$-$2.5$\pm$1.6&8.35$\pm$0.47&M2.3$^c$&\\
\hline
1571717&&19:24:10.52&+37:06:34.4&$-$0.9$\pm$1.9$^a$&2.1$\pm$0.9&9.03$\pm$0.36&A6.9$^b$&&9.2&5868&0.82\\
1571732&&19:24:11.27&+37:06:32.3&$-$0.8$\pm$3.7$^d$&2.4$\pm$3.7&8.95$\pm$0.4&F5.8$^b$&\\
\hline
1873370&&19:30:12.45&+37:20:19.3&2.6$\pm$2.1$^c$&$-$3.1$\pm$2.1&10.65$\pm$0.22&F9.1$^c$&&7.0&9450&0.84\\
1867358&&19:30:12.24&+37:20:25.9&2.3$\pm$2.0$^c$&$-$0.0$\pm$1.8&10.81$\pm$0.15&K7.6$^c$&32.879$\pm$0.735\\
\hline
\end{tabular}
\normalsize
\end{table}
\end{landscape}
\subsection{The planet-host fraction of wide binary stars}
To study how the presence of a wide binary companion affects the resulting planetary system we use our sample of wide binaries with companionship probabilities above 80\% and which had passed our giant removal criteria. This sample of 401 multiples contained 529 {\it Kepler} targets. Next we crossmatched each {\it Kepler} targets in the sample with the most recent list of {\it Kepler} Objects of Interest (KOI) published by \cite{Rowe2015}. These are divided into confirmed planets and planet candidates. If a binary system contained two {\it Kepler} target stars then we treated each {\it Kepler} target separately so a binary system where both {\it Kepler} targets were classified as a KOI we counted each of these as a separate planet-hosts. We restricted our sample to {\it Kepler} targets with $10<K_p<16$ and $0.5M_{\odot}<M<1.5M_{\odot}$ as there are few {\it Kepler} target stars outside these ranges (where we assume a mass from our SED fitting). Our separation greater than 6$\arcsec$ cut meant we largely sample binaries with projected separations above 3,000\,AU (see Figure~\ref{bin_sep}). After our restrictions on mass and $K_p$ magnitude we were left with 419 objects from which we identified 10 KOI candidate planet hosts and 3 KOI confirmed planet hosts\footnote{ We use the {\it NExtSci Disposition} for the candidate and confirmed planet host classifications} (with two additional candidate planet hosts with masses outside our $0.5M_{\odot}<M<1.5M_{\odot}$ range). Hence 3.1$\pm$0.9\% of our likely binary primaries are candidate or confirmed planet hosts. No binary system where both components were KOIs was identified.

\begin{figure}
\begin{center}
\includegraphics[scale=0.5]{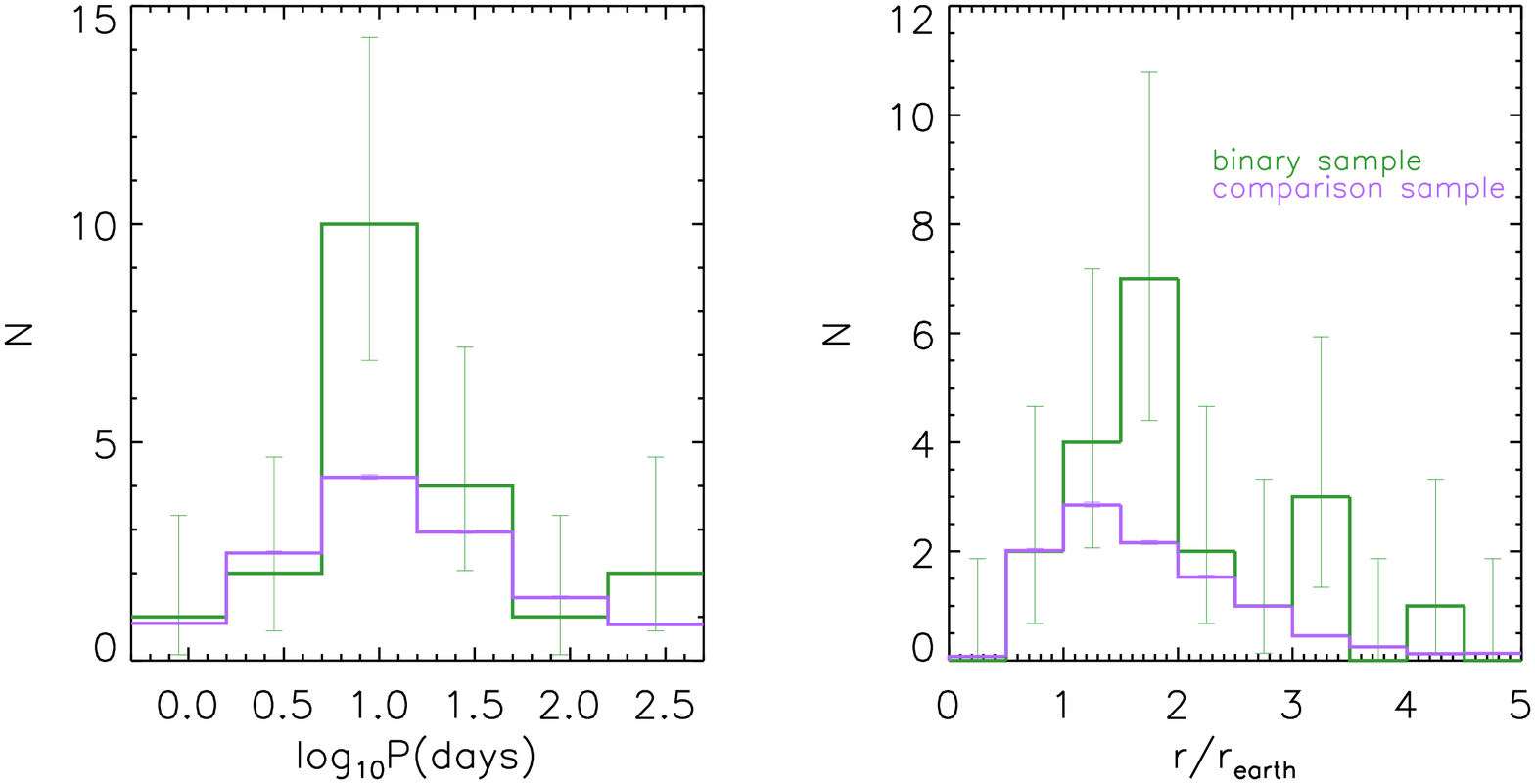}
\caption{\label{planet_prop} The properties of confirmed/candidate exoplanets around our binary components compared to a similar sample of {\it Kepler} targets. Our binaries exclude systems with separations below 6$\arcsec$.}
\end{center}
\end{figure}

We also constructed a comparison sample of {\it Kepler} targets. These were 159152 objects which passed our giant exclusion criteria and had $10<K_p<16$ and $0.5M_{\odot}<M<1.5M_{\odot}$. We found that our binary primaries were generally brighter and lower mass than the comparison sample of {\it Kepler} targets in general. This is likely due to our method selecting low-mass, nearby, high proper motion binaries preferentially as the number of coincident pairings with low-mass, nearby, high proper motion stars will be very low. To remedy this, we constructed 2D histograms for both our binary primaries and our comparison stars with bin sizes of 1\,mag in $K_p$ and 0.2$M_{\odot}$ in mass. We then used these to construct a weighting factor for each comparison star based on its $K_p$ magnitude and mass,

\begin{equation}
W(K_p,m)=\frac{N_{bin}(K_p,m)}{N_{comp}(K_p,m)}
\end{equation}
where $W$ is our weighting factor, $N_{comp}$ is the number of comparison stars in that bin and $N_{bin}$ is the number of binary primaries in that bin. The median bin in $N_{bin}$ contained 12 objects while the median bin in $N_{comp}$ contained 1804 objects.  To estimate the number of planet hosts in our sample we crossmatched our comparison sample with the list of KOIs and then summed the weighting factors to calculate how many planet hosts would be expected in a sample of comparison stars of identical size, magnitude distribution and mass distribution. This led us to a comparison KOI rate of 2.4$\pm$0.5\%. Hence we find no evidence for suppression or enhancement of the number of wide binary components which host exoplanets detectable by {\it Kepler}. We found that 1.0$\pm$0.5\% of our binaries had more than one Kepler confirmed or candidate planet while 0.5$\pm$0.1\% of our weighted sample did. This is again a statistically insignificant difference. We also examined the distributions of orbital period and planetary radius (see Figure~\ref{planet_prop}) finding no statistically significant differences between the candidate and confirmed exoplanets around our binary components and around a similar sample of {\it Kepler} targets.

Our work finds no statistically significant difference between the fraction of wide binary components which are identified as being Kepler Objects of Interest and a sample of Kepler targets of similar brightness and spectral type. While these KOIs are not all confirmed planet hosts, the fact that there is no significant difference indicates that in the regime where both our binary and comparison sample have detect planets (P$<$300 days) there is no evidence that a wide stellar companion with a separation wider than 3,000\,AU will have a large effect on planet occurrence. Note that we do not claim that our comparison sample only contains single stars as many will have unresolved companions or companions which are too faint. This sample relies on a small number of binaries containing 419 Kepler targets. The ongoing {\it Kepler K2} mission will drastically increase the number of stars in wide binaries probed for transiting exoplanets. This will increase the accuracy of the measurement of any possible effect from wide binarity. Additionally {\it K2} will focus more on M dwarfs, objects largely excluded from our analysis here. Finally it should be noted that our study does not disprove the theoretical hypothesis of \cite{Kaib2013} that wide binaries can affect the orbits of planets. This effect could still be extremely significant for planets in wider orbits similar to the gas giants in our own Solar System.

\section{Conclusions}
We have used Pan-STARRS\,1 data and archival datasets to calculate accurate proper motions for stars in the {\it Kepler} field. This catalogue will be available to the community online and will improve the characterisation of Kepler target stars (i.e. giant-dwarf separation, kinematic population membership etc.). By combining the proper motions with PS1-based SEDs to estimate the distance, we selected a sample wide binaries. To formalise this, we developed a statistical method to calculate true association probabilities for wide binaries including empirical estimates of the frequency of pairings of unrelated field stars. Our method was tested for a sample of bright stars showing that our likely binaries had low differences in radial velocity while our unlikely pairings had discrepant radial velocities. After excluding giant stars, we find 401 multiples containing at least one {\it Kepler} target which have binary companion probabilities of 0.8 or more and separations between six arcseconds and five arcminutes. We find no difference in the rate at which our wide binaries (typically wider than 3,000\,AU) are identified as confirmed or candidate exoplanet hosts than a comparison sample of {\it Kepler} targets of similar brightness and spectral type. An investigation of those binaries where both components have archival rotation periods shows that our binaries are roughly coeval indicating that our method is selecting true binaries.

\bibliography{ndeacon}\bibliographystyle{mn2e}
\section*{Acknowledgements}
The Pan-STARRS1 Surveys (PS1) have been made possible through contributions of the Institute for Astronomy, the University of Hawaii, the Pan-STARRS Project Office, the Max-Planck Society and its participating institutes, the Max Planck Institute for Astronomy, Heidelberg and the Max Planck Institute for Extraterrestrial Physics, Garching, The Johns Hopkins University, Durham University, the University of Edinburgh, Queen's University Belfast, the Harvard-Smithsonian Center for Astrophysics, the Las Cumbres Observatory Global Telescope Network Incorporated, the National Central University of Taiwan, the Space Telescope Science Institute, the National Aeronautics and Space Administration under Grant No. NNX08AR22G issued through the Planetary Science Division of the NASA Science Mission Directorate, the National Science Foundation under Grant No. AST-1238877, the University of Maryland, and Eotvos Lorand University (ELTE). We used public UK Infrared Telescope data taken from program U/09A/2 (PI Lucas) this used the UKIRT Wide Field Camera (WFCAM; \citealt{Casali2007}) and a photometric system described in \cite{Hewett2006}. The pipeline processing and science archive are described in \cite{Irwin2004} and \cite{Hambly2008}. The United Kingdom Infrared Telescope was operated by the Joint Astronomy Centre on behalf of the Science and Technology Facilities Council of the U.K. The authors wish to recognize and acknowledge the very significant cultural role and reverence that the summit of Mauna Kea has always had within the indigenous Hawaiian community. We are most fortunate to have the opportunity to conduct observations from this mountain. The authors would like to thank Eddie Schlafly for discussions on correlated errors.
 \appendix
 \section{Determining companionship probabilities}
 \label{probmaths}
 In order to define a formal companionship probability, we define the following likelihood,
 \begin{equation}
\phi=\phi_c(\Delta \mu,\mu,\Delta D,D,s,m)+\phi_f(\mu,D,s,m)
 \end{equation}
 where $\phi_c$ and $\phi_f$ are the companion and coincident field distributions and are functions for the separation ($s$), proper motion difference ($\Delta \mu$), proper motion ($\mu$) and distance modulus difference ($\Delta D$) between any particular pair and the mass of the secondary component ($m$). The probability of a pair being real rather than a coincident pairing of unrelated field stars is
 \begin{equation}
 p=\frac{\phi_c}{\phi_c+\phi_f}
 \end{equation}
 First let us examine the companion distribution,
 \begin{equation}
 \label{phic1}
 \phi_c = f_{bin} f_{det} \times \left[ \frac{e^{-\Delta \mu^2/2\sigma_{\mu}^2}}{2\pi \sigma_{\mu}^2}\right]\times\left[\frac{e^{-\Delta D^2/2\sigma_D^2}}{\sqrt{2\pi}\sigma_D}\right]\times\left[ c_s \left( sd \right)^{-x}\right]\times \left[ c_m f(m) \right]
 \end{equation}
 Where $\Delta \mu$ is the proper motion difference, $\sigma_{\mu}$ is the quadrature sum of the proper motion errors, $\Delta D$ is the difference is distance moduli, $\sigma_{D}$ is the quadrature sum of the distance modulus errors, $s$ is the separation on the sky, $m$ is the mass of the secondary and $d$ is the calculated distance to the primary. The last two sets of brackets on the right each contain normalisation terms. We assume a power-law in physical separation but this method could be adapted to any underlying separation distribution or binary fraction. The separation distribution is normalised by $c_s$,
 \begin{equation}
 \label{cs1}
 c_s = 1/ \int_{s_1}^{s_2} \left( sd \right)^{-x}ds
 \end{equation}
Where $s_1$ and $s_2$ are the inner and outer search radii of our survey. There are two free parameters to choose to complete this system, the first is the exponent $x$ which defines the separation distribution. The other is the binary fraction $f_{bin}$ defined as the fraction of stars which have a companion in the region in which our separation distribution is defined. Next we must consider our observational selection effect caused by our choice of an inner and outer search separation. This is written in Equation~\ref{phic1} as $f_{det}$ and can be defined as,
\begin{equation}
\label{fdet}
f_{det} = d \int_{s_1}^{s_2} \left( sd \right)^{-x}ds /  \int_{r_{min}}^{r_{max}} r^{-x}dr
\end{equation}
Where $r$ is the physical separation in AU (defined as $sd$) and our separation distribution is defined in the region $r_{min}$ to $r_{max}$. Note the numerator of Equation~\ref{fdet} is equal to the denominator of Equation~\ref{cs1}. Hence we can condense these two factors so that,
\begin{equation}
c'_s=c_s \times f_{det} = d/ \int_{r_{min}}^{r_{max}} r^{-x}dr
\end{equation}
The final term in Equation~\ref{phic1} is the mass distribution of likely companions ($f(m)$) with the normalisation $c_m$. For this study we assume a flat mass ratio distribution. Hence as the mass of the secondary cannot be greater than the mass of the primary, the normalisation $c_m$ becomes 1/$m_p$ where $m_p$ is the mass of the primary (assumed to be the earlier-type star in the pair).

 \begin{equation}
 \label{phic2}
 \phi_c = f_{bin} \times \left[ \frac{e^{-\Delta \mu^2/2\sigma_{\mu}^2}}{2\pi \sigma_{\mu}^2}\right]\times\left[\frac{e^{-\Delta D^2/2\sigma_D^2}}{\sqrt{2\pi}\sigma_D}\right]\times\left[ c'_s \left(sd \right)^{-x}\right]\times\left[ \frac{1}{m_p}\right]
 \end{equation}
Next let us turn to the field distribution. The Galactic field star population is complex with the proper motion distribution particularly hard to parameterise over a wide range of proper motions. Hence we adapt the positional offset test of \cite{Lepine2007} to measure the likely contamination rate. Defining,
\begin{equation}
\phi_f = c_f 2 \pi s 
\end{equation}
where $c_f$ is a the density of stars of similar absolute proper motion and has units of stars per (mas/yr.)$^2$ per magnitude per sq. arcsecond per solar mass. The factor $c_f$ is defined empirically by offsetting the positions of primary stars in our sample and repairing them to produce entirely confident pairs \citep{Lepine2007}. We do this and (for each of our actual pairs) measure the density of coincident pairs at the particular magnitude difference, proper motion difference and mass ratio of the actual pair.

\section{Empirical SEDs in the Pan-STARRS\,1 photometric system}
\label{sed_sect}
In order to calculate accurate photometric distances for stars in the {\it Kepler} field we require a series of empirical stellar Spectral Energy Distributions (SEDs). There are multiple sources of empirical stellar SEDs with the most commonly used being \cite{Kraus2007} for the SDSS/2MASS systems and \cite{Pecaut2013} for the Cousins/2MASS/WISE systems. There are no empirical SEDs in the Pan-STARRS\,1 photometric system defined by \cite{Tonry2012} over the required range of spectral types (A--M). \cite{Dupuy2009} present synthetic colours for a range of spectral types based on initial characterisation of the throughput of the Pan-STARRS\,1 photometric system. \cite{Aller2013} present a series of different SED templates for M dwarfs. These include templates for M and early L dwarfs in the star forming region Upper Sco and early M dwarfs based on members of the Praesepe and Coma Ber clusters. Finally \cite{Aller2013} presents individual M9 and L0 dwarfs to be used templates for field stars. Note that \cite{Aller2013} used photometry from a previous processing version to our work so there may be subtle differences between the photometric systems.

We began by taking the large collection of known dwarfs in the {\it Kepler} field with spectrally determined effective temperature from \cite{Huber2014} and references therein. We converted their effective temperatures to spectral type using the dwarf effective temperature scale of \cite{Pecaut2013}. For M~dwarfs from the work of \cite{Martin2013} we note that the effective temperatures in this work are derived from SED fitting so instead we convert the objects' determined spectral types to effective temperatures using the \cite{Pecaut2013} scale. We supplement the objects from the \cite{Huber2014} collection with late M~dwarfs drawn from the compilation of \cite{Faherty2009} as these objects typically have high proper motions we corrected their positions to an epoch 2010.0 for comparing to Pan-STARRS\,1 data. We also added M dwarfs from the sample of \cite{West2008} taking only bright ($z_{sdss}<16$) and low extinction ($A_g<0.05$) objects. We drew Pan-STARRS\,1 photometry from the nightly science database and 2MASS data from the 2MASS catalogue using pairing radii of 3 arcseconds. Objects were only incorporated into the fit in a particular PS1 band if they had a magnitude error less than 0.05, more than 2 detections, were not saturated in that band, were not extended in that band and had been subject to the "ubercal" calibration process \citep{Schlafly2012}.  

We note that there are few objects in the \cite{Huber2014} catalogue hotter than 6000\,K which are not saturated in at least one Pan-STARRS\,1 band. Hence we supplemented our objects with synthetic colours calculated from \cite{Pickles1998} and Dwarf Archives optical spectra and near-IR spectra from \cite{Cushing2005} and \cite{Rayner2009}. The Pickles spectra overlapped with those from Rayner over the $y_{P1}$ band so we calculated mean optical colours relative to $y_{P1}$ in the AB system for each spectral type. We then calculated the colours relative to $J_{2MASS}$ by calculating $y_{P1}-J_{2MASS}$ from the Rayner spectra, adding a term to take into account that $y_{P1}$ is in the AB system and $J_{2MASS}$ is in the Vega system. 

\begin{figure}
\begin{center}

\includegraphics[scale=0.7]{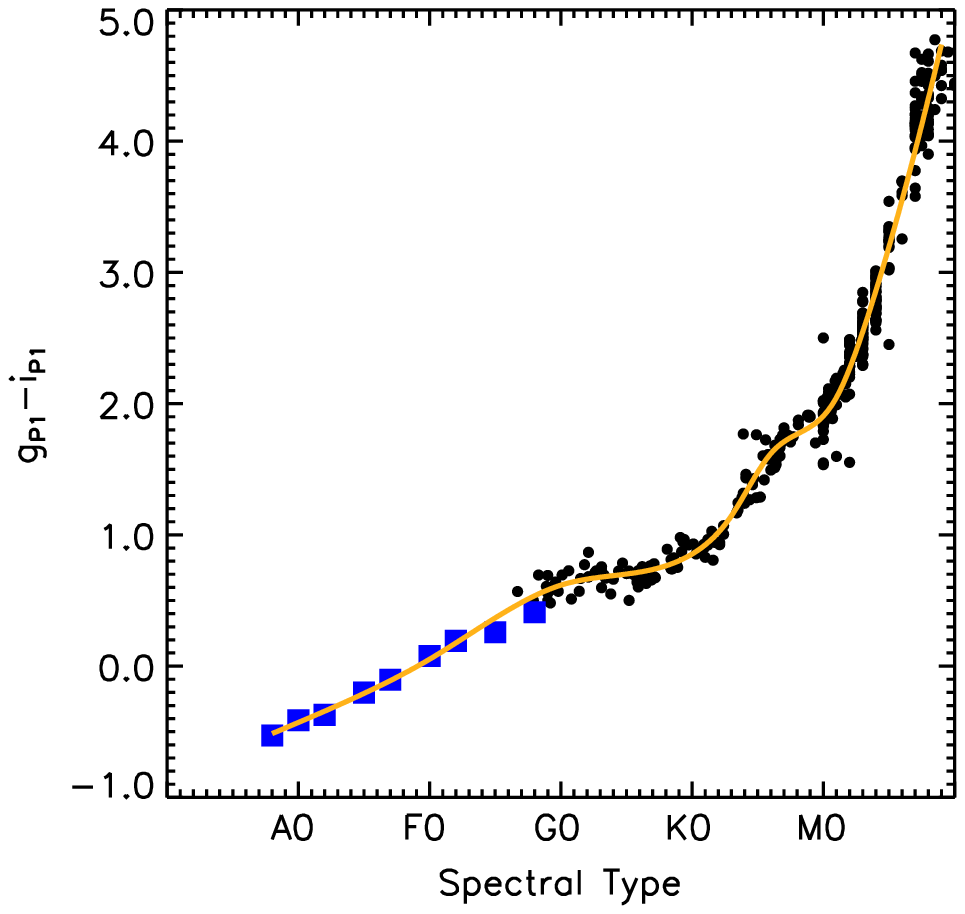}
\caption{\label{spt_gi} The spectral type to $g_{p1}-i_{P1}$ transformation used to calibrate our SED templates. The black points are observed stars of known temperature/spectral type and the blue squares are synthetic photometry derived from the spectra of \protect\cite{Pickles1998} and \protect\cite{Rayner2009}. The yellow line is a cubic spline fit to the data.}
\end{center}
\end{figure}

\begin{figure}
\begin{center}
\includegraphics[scale=0.7]{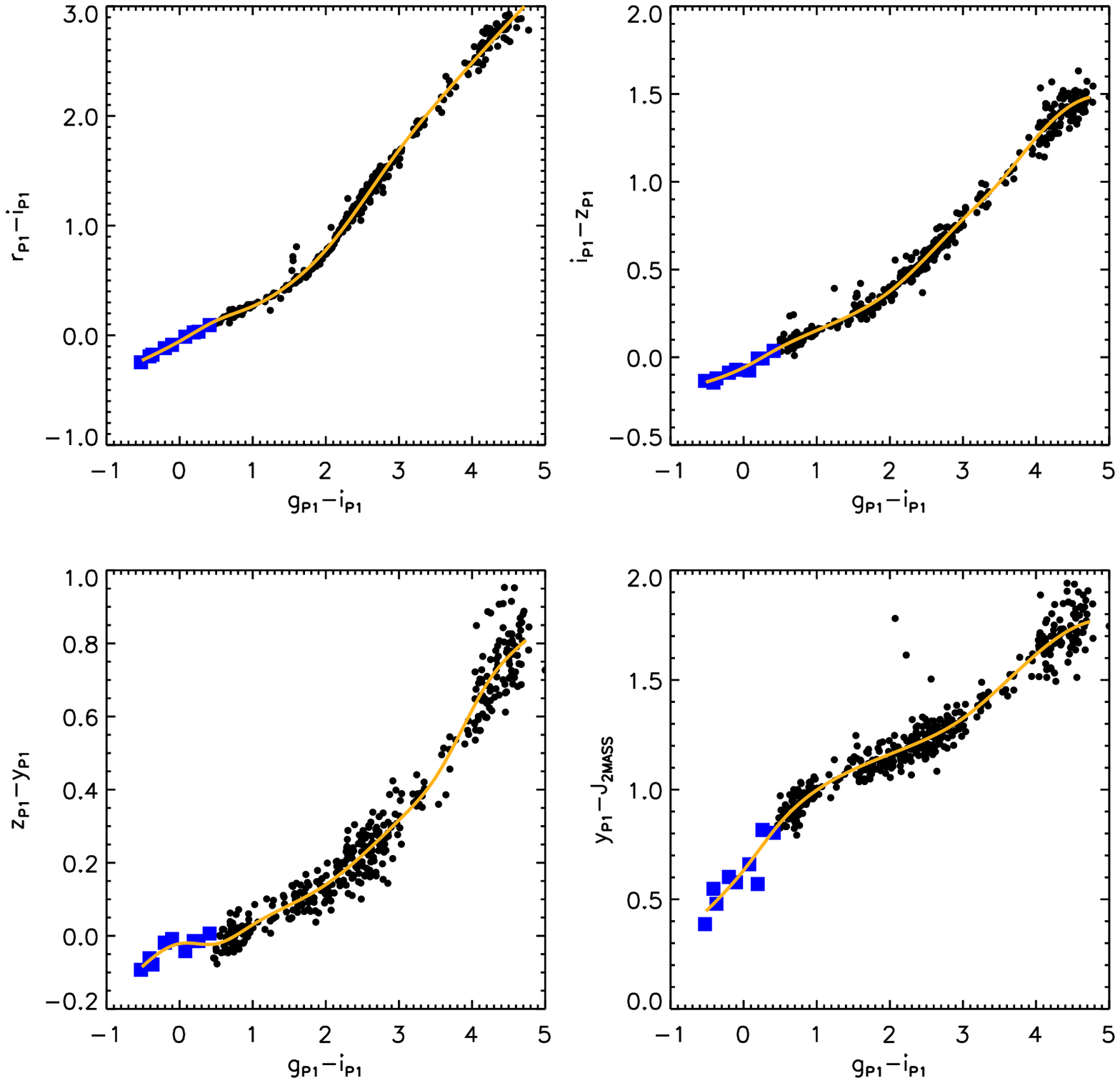}
\caption{\label{spline_test} The transformation between $g_{p1}-i_{P1}$ and various other colours used to calibrate our SED templates. The black points are observed stars of known temperature/spectral type and the blue squares are synthetic photometry derived from the spectra of \protect\cite{Pickles1998} and \protect\cite{Rayner2009}. The yellow line is a cubic spline fit to the data.}
\end{center}
\end{figure}

For the Pan-STARRS\,1 bands we developed the following mechanism to calculate the magnitudes for each spectral type in each band. We began by fitting our $g_{P1}-i_{P1}$ colours with a spline to produce a smooth distribution relative to spectral type. For spectral type we used ten subclasses per spectral classification.This is a departure from the traditional picture where there are no K6, K8 and K9 dwarfs. However we use the effective temperature scale of \cite{Pecaut2013} who include these missing spectral types in their temperature scale. Our $g_{P1}-i_{P1}$ to spectral type fit is shown in Figure~\ref{spt_gi}. We then produced splines for relations between $g_{P1}-i_{P1}$ and other colours; $r_{P1}-i_{P1}$, $i_{P1}-z_{P1}$, $z_{P1}-y_{P1}$ and $y_{P1}-J_{2MASS}$. These fits are shown in Figure~\ref{spline_test}. We then used these colours at each $g_{P1}-i_{P1}$ (and hence spectral type) to produce colours relative to the 2MASS $J$-band magnitude. 

These were then combined with the bolometric magnitudes from \cite{Kraus2007} and the bolometric correction and $V-J$ colours of \cite{Pecaut2013}. for objects later than M7 we used the spectral type -- $J$-band absolute magnitude relations of \cite{Dupuy2012}. For near-infrared colours, we used $J_{2MASS}-H_{2MASS}$ and $J_{2MASS} - K_{s\,2MASS}$ colours derived from the bolometric corrections of \cite{Pecaut2013}. We also synthesised $J_{2MASS}-J_{MKO}$, $H_{2MASS}-H_{MKO}$ and $K_{s\,2MASS}-K_{MKO}$ from the Rayner spectra to derive the MKO magnitudes \citep{Tokunaga2002} of all our spectral types earlier than M7 and used the quoted \cite{Dupuy2012} spectral type -- absolute magnitude relations for spectral types M7 and later. Our input sample included a handful of M9.5 and L0 objects to anchor our fits at the cool end. We do not quote an L0 template. Our final SED templates for each subtype between B9 and M9 are shown in Table~\ref{sed_templates}.

\clearpage
%\begin{landscape}
\begin{table}
\scriptsize
\caption{\label{sed_templates} The SED templates in the Pan-STARRS\,1 and MKO photometric systems used in this work. The bolometric magnitudes and masses come from the compilation of \protect\cite{Kraus2007} and the effective temperatures are from the dwarf effective temperature scale is the dwarf scale from \protect\cite{Pecaut2013}. Our templates also used the 2MASS colours from \protect\cite{Pecaut2013} before M6 and from \protect\cite{Dupuy2012}. The reader should note that \protect\cite{Pecaut2013} and \protect\cite{Dupuy2012} also include WISE colours.}
\begin{tabular}{crrrrrrrrrrr}
\hline
{\bf SpT}&{\bf $g_{P1}$}&{\bf $r_{P1}$}&{\bf $i_{P1}$}&{\bf $z_{P1}$}&{\bf $y_{P1}$}&{\bf $J_{MKO}$}&{\bf $H_{MKO}$}&{\bf $K_{MKO}$}&{\bf $M_{bol}$}&{\bf $T_{eff}$}&{\bf mass}\\
&{\bf (mag.)}&{\bf (mag.)}&{\bf (mag.)}&{\bf (mag.)}&{\bf (mag.)}&{\bf (mag.)}&{\bf (mag.)}&{\bf (mag.)}&{\bf (mag.)}&{\bf K}&{\bf $M_{\odot}$}\\
\hline
B8.0V&$-$0.31&$-$0.03&0.2&0.34&0.43&0.02&0.06&0.16&$-$1.0&12500&3.8\\
B9.0V&$-$0.05&0.21&0.42&0.56&0.64&0.21&0.22&0.29&$-$0.35&10700&3.35\\
A0.0V&0.34&0.57&0.77&0.9&0.97&0.51&0.51&0.54&0.3&9700&2.9\\
A1.0V&0.67&0.87&1.06&1.18&1.24&0.78&0.76&0.79&0.7&9200&2.65\\
A2.0V&1.02&1.19&1.36&1.48&1.54&1.07&1.03&1.06&1.1&8840&2.4\\
A3.0V&1.24&1.38&1.54&1.65&1.7&1.21&1.17&1.2&1.32&8550&2.267\\
A4.0V&1.41&1.52&1.67&1.77&1.81&1.3&1.24&1.28&1.53&8270&2.133\\
A5.0V&1.65&1.73&1.86&1.96&2.0&1.46&1.41&1.44&1.75&8080&2.0\\
A6.0V&1.87&1.92&2.03&2.12&2.15&1.6&1.54&1.58&1.92&8000&1.9\\
A7.0V&2.02&2.03&2.13&2.21&2.23&1.67&1.59&1.63&2.08&7800&1.8\\
A8.0V&2.2&2.18&2.26&2.33&2.36&1.77&1.67&1.71&2.26&7500&1.733\\
A9.0V&2.46&2.41&2.46&2.53&2.55&1.93&1.83&1.86&2.43&7440&1.667\\
F0.0V&2.67&2.58&2.62&2.67&2.69&2.04&1.92&1.95&2.61&7200&1.6\\
F1.0V&2.84&2.72&2.73&2.76&2.78&2.12&1.97&1.98&2.75&7030&1.55\\
F2.0V&3.02&2.86&2.84&2.86&2.88&2.2&2.02&2.02&2.89&6810&1.5\\
F3.0V&3.35&3.15&3.11&3.11&3.14&2.42&2.23&2.23&3.13&6720&1.417\\
F4.0V&3.66&3.42&3.36&3.35&3.37&2.63&2.42&2.43&3.37&6640&1.333\\
F5.0V&3.96&3.68&3.59&3.57&3.59&2.83&2.6&2.61&3.61&6510&1.25\\
F6.0V&4.2&3.89&3.78&3.74&3.76&2.96&2.71&2.72&3.82&6340&1.223\\
F7.0V&4.48&4.12&3.99&3.94&3.96&3.13&2.88&2.88&4.03&6240&1.197\\
F8.0V&4.75&4.36&4.21&4.15&4.17&3.31&3.05&3.06&4.24&6150&1.17\\
F9.0V&4.91&4.49&4.33&4.26&4.27&3.4&3.13&3.13&4.35&6040&1.14\\
G0.0V&5.04&4.59&4.42&4.34&4.35&3.47&3.19&3.18&4.47&5920&1.11\\
G1.0V&5.13&4.66&4.49&4.4&4.41&3.52&3.22&3.21&4.53&5880&1.085\\
G2.0V&5.16&4.68&4.5&4.41&4.42&3.52&3.19&3.18&4.6&5770&1.06\\
G3.0V&5.27&4.78&4.59&4.5&4.51&3.61&3.27&3.26&4.7&5720&1.053\\
G4.0V&5.38&4.87&4.69&4.59&4.6&3.69&3.35&3.34&4.79&5680&1.047\\
G5.0V&5.48&4.97&4.78&4.68&4.69&3.78&3.44&3.42&4.89&5660&1.04\\
G6.0V&5.62&5.1&4.91&4.81&4.81&3.9&3.54&3.52&5.03&5590&1.02\\
G7.0V&5.79&5.25&5.05&4.95&4.95&4.03&3.67&3.65&5.16&5530&1.0\\
G8.0V&5.94&5.38&5.17&5.06&5.06&4.13&3.76&3.74&5.3&5490&0.98\\
G9.0V&6.17&5.58&5.37&5.25&5.24&4.3&3.91&3.89&5.49&5340&0.94\\
K0.0V&6.39&5.76&5.54&5.41&5.4&4.43&4.03&4.0&5.69&5280&0.9\\
K1.0V&6.67&5.99&5.75&5.61&5.59&4.61&4.18&4.15&5.89&5170&0.86\\
K2.0V&6.96&6.21&5.94&5.78&5.75&4.75&4.29&4.26&6.08&5040&0.82\\
K3.0V&7.32&6.48&6.17&5.99&5.94&4.92&4.41&4.37&6.32&4840&0.785\\
K4.0V&7.68&6.75&6.37&6.16&6.1&5.04&4.49&4.43&6.55&4620&0.75\\
K5.0V&8.09&7.07&6.62&6.37&6.29&5.2&4.63&4.56&6.68&4450&0.7\\
K6.0V&8.31&7.22&6.69&6.42&6.32&5.22&4.59&4.49&6.78&4200&0.665\\
K7.0V&8.51&7.38&6.8&6.5&6.4&5.29&4.64&4.52&6.89&4050&0.63\\
K8.0V&8.81&7.66&7.04&6.74&6.62&5.5&4.85&4.72&7.13&3970&0.617\\
K9.0V&9.11&7.94&7.29&6.96&6.84&5.71&5.07&4.92&7.36&3880&0.603\\
M0.0V&9.46&8.27&7.56&7.21&7.08&5.93&5.3&5.15&7.6&3850&0.59\\
M1.0V&9.99&8.76&7.94&7.55&7.4&6.24&5.64&5.45&7.97&3680&0.54\\
M2.0V&10.8&9.54&8.54&8.06&7.88&6.68&6.07&5.87&8.44&3550&0.42\\
M3.0V&11.9&10.61&9.35&8.76&8.53&7.28&6.7&6.48&9.09&3400&0.29\\
M4.0V&13.27&11.97&10.41&9.68&9.39&8.06&7.53&7.26&9.92&3200&0.2\\
M5.0V&15.03&13.69&11.84&10.97&10.61&9.19&8.65&8.35&11.01&3050&0.15\\
M6.0V&16.58&15.18&13.03&12.01&11.56&10.04&9.46&9.12&12.06&2800&0.12\\
M7.0V&17.88&16.38&13.95&12.73&12.14&10.56&10.01&9.57&12.7&2650&0.11\\
M8.0V&19.05&17.45&14.73&13.34&12.62&10.94&10.36&9.87&13.13&2570&0.102\\
M9.0V&20.05&18.33&15.31&13.83&13.02&11.21&10.59&10.09&13.43&2450&0.088\\
\hline
\end{tabular}
\normalsize
\end{table}
%\end{landscape}

\section{Pan-STARRS\,1 to {\it Kepler} photometric transformations}
\label{kp_sect}
\begin{figure}
\begin{center}
\begin{tabular}{cc}

\includegraphics[scale=0.5]{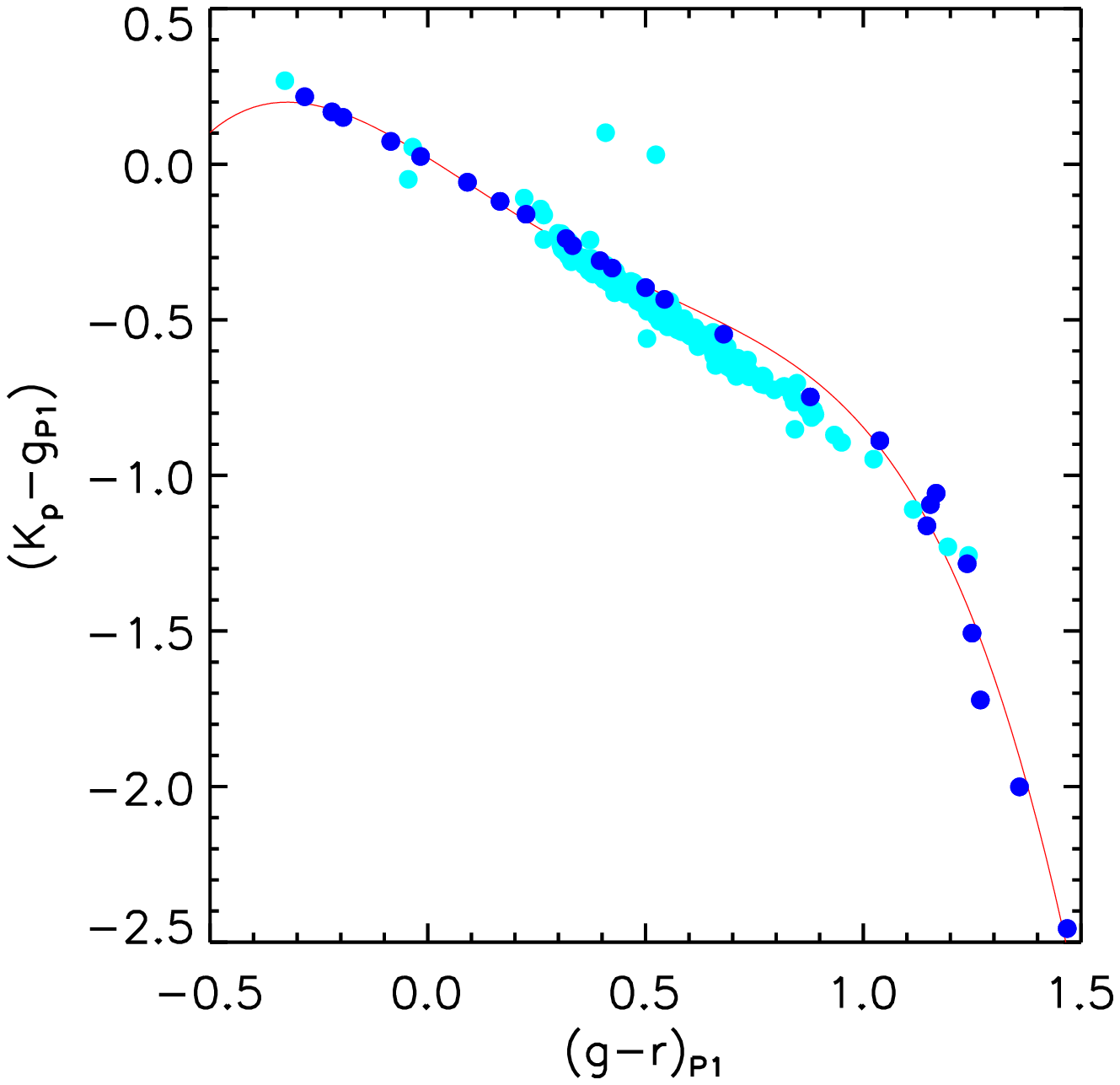}&
 \includegraphics[scale=0.5]{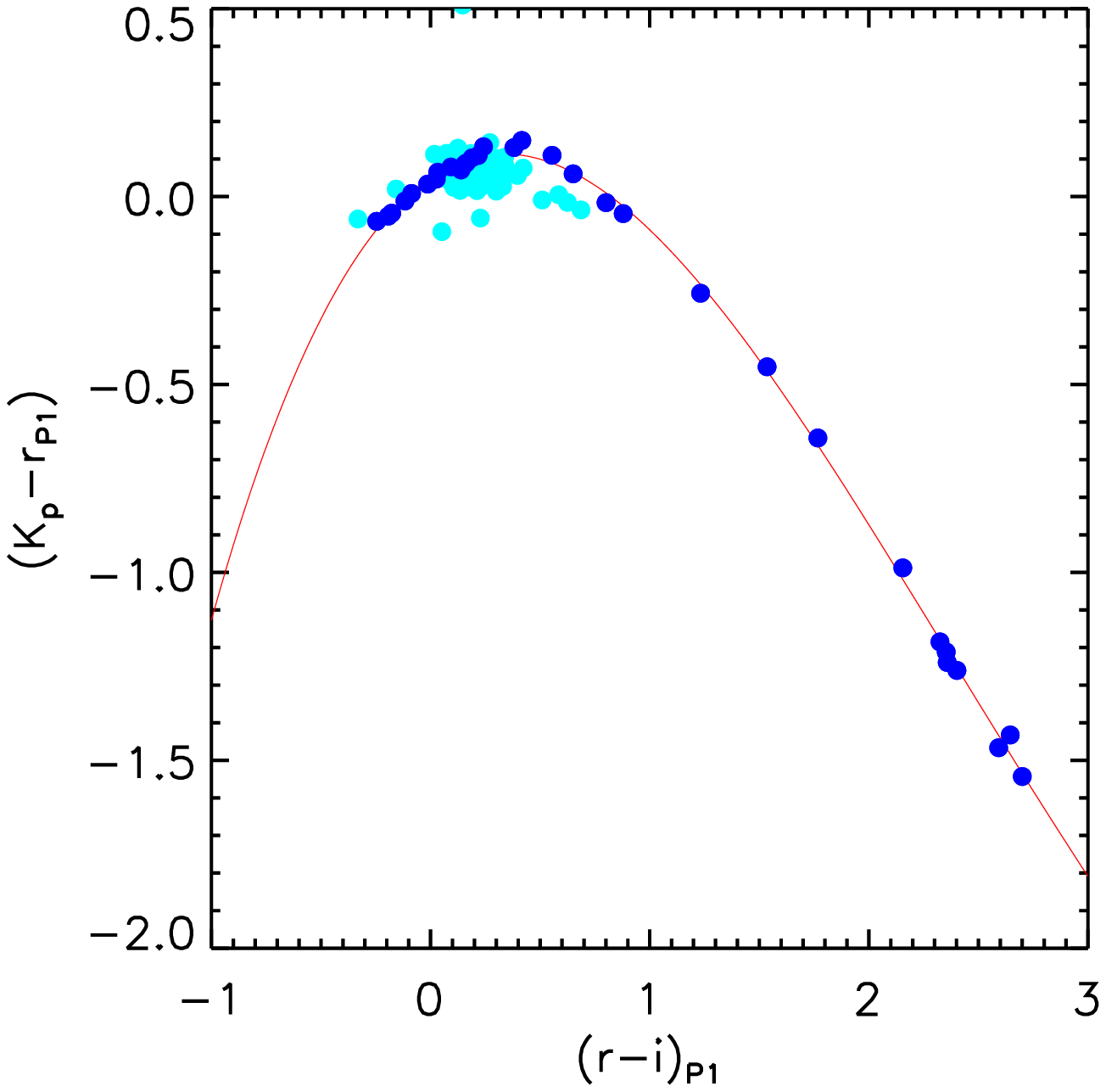}\\
\includegraphics[scale=0.5]{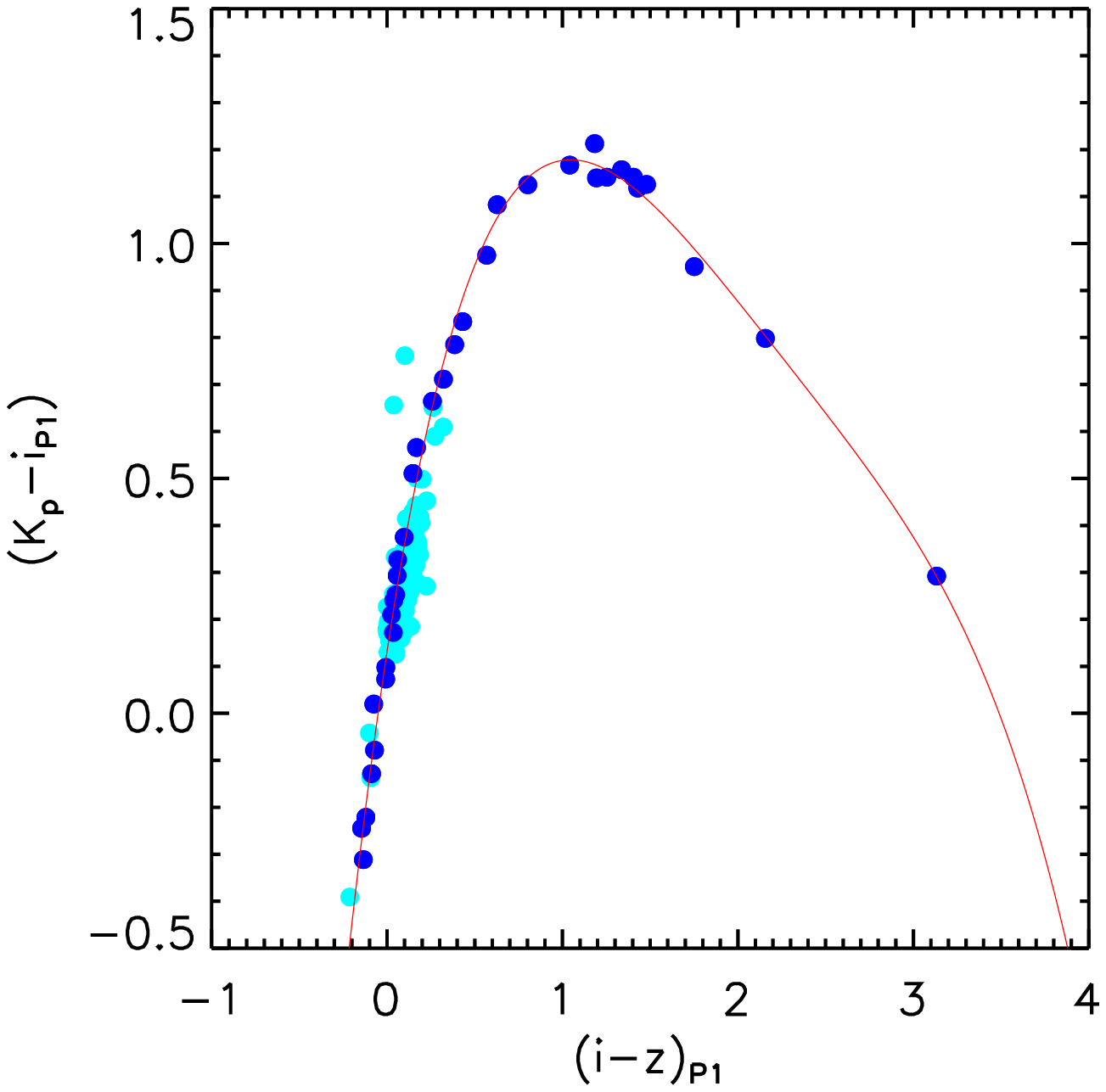}&
 \includegraphics[scale=0.5]{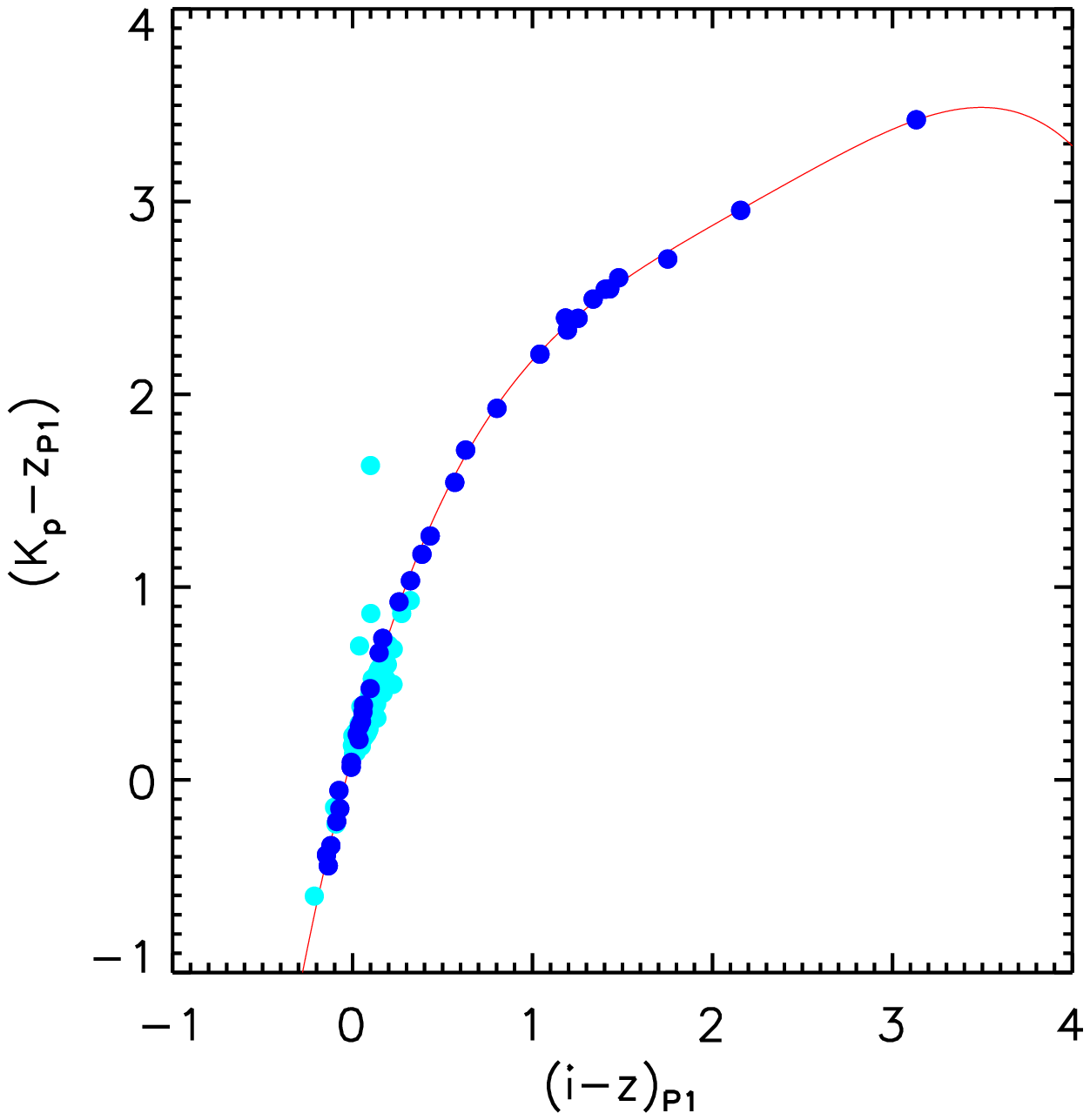}\\
\end{tabular}
\caption{\label{Kp_colours} Synthesised colors for Pan-STARRS\,1 to {\it Kepler} magnitudes for empirical spectra from \protect\cite{Pickles1998}, Dwarf Archives and \protect\cite{Leggett2010} (blue points). The red line shows the fifth degree polynomial fits listed in Table~\ref{Kp_colour_fit}. The aqua points are from the synthesised $K_p$ magnitudes for the standards of \protect\cite{Brown2011} and the observed Pan-STARRS\,1 magnitudes of these objects.}
\end{center}
\end{figure}

Pan-STARRS\,1 provides CCD photometry for both the original {\it Kepler} field and the ten K2 mission fields. {\it Kepler} magnitudes are derived from a number of sources including Tycho \cite{Hog2000}, photographic plates and $g$, $r$, $i$ photometry in the {\it SDSS} system from the {\it Kepler} Spectroscopic Classification Program \cite{Brown2011}. There are also additional datasets such as the {\it Kepler} INT Survey (KIS; \citealt{Greiss2012}). In order to enable the calculation of {\it Kepler} magnitudes from Pan-STARRS\,1 data we synthesised {\it Kepler} magnitudes for a series of stars using the dwarf spectral sequence of \cite{Pickles1998} for types B9 to M6. We supplemented these with seven late M dwarfs from the Dwarf Archives spectral library \footnote{http://spider.ipac.caltech.edu/staff/davy/ARCHIVE/index.shtml} with sufficient spectral coverage and three L and T dwarfs from the compilation of \cite{Leggett2010}. Note for the \cite{Leggett2010} spectra we included objects with wavelength coverage until 4000\AA.~This leaves the 3800--4000\AA~region with no data in the {\it Kepler} filter bandpass, however for objects this cool the emission in this region will be negligible. We synthesised colours using the {\it Kepler} bandpass \footnote{$http://keplergo.arc.nasa.gov/kepler_response_hires1.txt$} and the Pan-STARRS\,1 filter set \citep{Tonry2012}. All colours were calculated in the AB system. Our colours are shown in Figure~\ref{Kp_colours} along with fifth degree polynomial fits. These coefficients of these fits are shown in Table~\ref{Kp_colour_fit}. Figure~\ref{Kp_colours} also shows the synthesised {\it Kepler} magnitudes for the {\it Kepler} standards listed in \cite{Brown2011} plotted against their observed Pan-STARRS1 magnitudes. Except for two outliers which lie near the saturation point for Pan-STARRS\,1, there is clearly fairly good agreement between our synthesised {\it Kepler}-Pan-STARRS\,1 and the Brown synthesised {\it Kepler}-real Pan-STARRS\,1 colors. The small offsets are likely due to the differences in the ways the {\it Kepler} magnitudes were synthesised between the two sources.

\begin{table}
\scriptsize
\caption{\label{Kp_colour_fit} Color transformations from Pan-STARRS\,1 filters to {\it Kepler} magnitudes. These take the form $y=\sum_{i=0}^{4} c_i x^i$ with the applicable colour range for each fit listed as $x_{range}$}
\begin{tabular}{llcccccc}
\hline
{\bf $y_{color}$} &
{\bf $x_{color}$} &
{\bf $c_0$} &
{\bf $c_1$} &
{\bf $c_2$} &
{\bf $c_3$} &
{\bf $c_4$}&
{\bf $x_{range}$}\\
\hline
$K_p$-$g_P1$&$g_P1$-$r_P1$&0.0208986&$-$0.871090&$-$0.381325&1.50797&$-$1.12176&$-$0.28--1.46\\
$K_p$-$r_P1$&$r_P1$-$i_P1$&0.0502748&0.393252&$-$0.650769&0.126130&$-$0.00725137&$-$0.24--2.70\\
$K_p$-$i_P1$&$i_P1$-$z_P1$&0.131700&2.48210&$-$1.94587&0.572845&$-$0.0636664&$-$0.15--3.13\\
$K_p$-$z_P1$&$i_P1$-$z_P1$&0.131700&3.48210&$-$1.94587&0.572845&$-$0.0636664&$-$0.15--3.13\\
\hline
\normalsize
\end{tabular}
\end{table}
\label{lastpage}
\end{document}